\documentclass{aa}

\usepackage{amssymb}
\usepackage{wasysym}
\usepackage{graphicx}
\usepackage[varg]{txfonts}
\usepackage{multirow}
\usepackage{natbib}
\usepackage{caption}
\usepackage{subcaption}

\bibpunct{(}{)}{;}{a}{}{,} 

\begin{document}
        
        \title{Detection of new eruptions in the Magellanic Clouds LBVs R\,40 and R\,110\thanks{Based on observations with the 0.6\,m telescope at Pico dos Dias Observatory (Brazil) and MPG/ESO 2.2-m telescope at the European Southern Observatory (La Silla, Chile) under the Prog. IDs: 094.A-9029(D), 096.A-9039(A), and 098.A-9039(C), and under the agreements ESO-Observatório Nacional/MCTIC and MPI-Observatório Nacional/MCTIC, Prog. IDs.: 076.D-0609(A) and 096.A-9030(A).}}

        \author{J. C. N. Campagnolo\inst{1} \and
                M. Borges Fernandes\inst{1} \and
                N. A. Drake\inst{1,2} \and
                M. Kraus\inst{3,4} \and
                C. A. Guerrero\inst{1} \and
                C. B. Pereira\inst{1}}
        
        \institute{Observat\'{o}rio Nacional,
                Rua General Jos\'{e} Cristino, 77, S\~{a}o Crist\'{o}v\~{a}o, Rio de Janeiro, 20921-400, Brazil\\
                \email{juliocampagnolo@on.br}
        \and
        Laboratory of Observational Astrophysics, Saint Petersburg State University, Universitetski pr. 28, 198504 Saint Petersburg, Russia
        \and
        Astronomick\'y \'ustav, Akademie v\v{e}d \v{C}esk\'e republiky, Fri\v{c}ova 298, 251\,65 Ond\v{r}ejov, Czech Republic
        \and
        Tartu Observatory, 61602 T{\~o}ravere, Tartumaa, Estonia
        }
        
        \date{Received 18 August 2017; accepted 06 November 2017}
        
\abstract
{
We performed a spectroscopic and photometric analysis to study new eruptions in two luminous blue variables (LBVs) in the Magellanic Clouds. We detected a strong new eruption in the LBV R40 that reached $V \sim 9.2$ in 2016, which is around $1.3$ mag brighter than the minimum registered in 1985. During this new eruption, the star changed from an A-type to a late F-type spectrum. Based on photometric and spectroscopic empirical calibrations and synthetic spectral modeling, we determine that R\,40 reached $T_{\mathrm{eff}} = 5800-6300$~K during this new eruption. This object is thereby probably one of the coolest identified LBVs. We could also identify an enrichment of nitrogen and r- and s-process elements. We detected a weak eruption in the LBV R 110 with a maximum of $V \sim 9.9$ mag in 2011, that is, around $1.0$ mag brighter than in the quiescent phase. On the other hand, this new eruption is about $0.2$ mag fainter than the first eruption detected in 1990, but the temperature did not decrease below 8500 K. Spitzer spectra show indications of cool dust in the circumstellar environment of both stars, but no hot or warm dust was present, except by the probable presence of PAHs in R\,110. We also discuss a possible post-red supergiant nature for both stars.
}

  
  \maketitle

\section{Introduction}

Massive stars experience some not well understood phases in their final evolution, such as the luminous blue variable (LBV) phase. Only $\sim$40 LBVs are actually known in the Galaxy and in the galaxies of the Local Group \citep{2012ASSL..384..221V}, indicating a short phase in the stellar life of $\sim$25000 years \citep{1994PASP..106.1025H}. These stars, also named as S Dor Variables \citep{2001A&A...366..508V}, are mainly characterized by episodes of strong mass loss (eruptions), resulting in irregular photometric and spectroscopic variabilities.

The role of LBVs in the stellar evolution of massive stars still remains an open question. In the temperature and luminosity range in which LBVs are expected, other types of objects are also seen, such as blue supergiants (BSG) and late-type Wolf-Rayet (WNL) stars, rendering it difficult to identify each phase based purely on its physical parameters, even obtained from a deep analysis of the chemical composition derived from the stellar evolution models \citep{2012A&A...538L...8G}. For the most massive stars, with $M_\mathrm{ZAMS} > 40\,M_\odot$, it is believed that the LBV phase is a transition phase from the main sequence -- or possibly BSG phase -- to Wolf-Rayet, that is, the star transits from the end of H-core burning to the beginning of He-core burning phase \citep{1983A&A...120..113M, 1999PhDT.........3V, 2013EAS....60...31E, 2014A&A...564A..30G}. For lower masses ($20 \leq M_\mathrm{ZAMS} \leq 25\,M_\odot$), \citet{2014A&A...564A..30G} cite that the LBV phase can only occur in advanced stages of evolution, during the burning of He in the core of the star, associated with a post-red supergiant (RSG) phase and a pre- supernova (SN) stage \citep{2013A&A...558A.131G}. For stars with $M_\mathrm{ZAMS} < 20\,M_\odot$, it is not expected that an LBV phase will occur during their evolution.

According to \citet{1999PhDT.........3V}, there are three types of variations that can be detected in LBVs: \textit{microvariations}, with $\Delta V < 0.3$ mag on timescales from weeks to months, possibly associated with non-radial pulsations; \textit{moderate variations}, with $1 < \Delta V < 2$ mag and periodicity from years to decades, possibly associated with radial pulsations; and \textit{eruptions}, with $\Delta V > 2$ mag, without a precise periodicity and associated with the strongest mass-loss episodes, wherein the most extreme cases are also called \textit{great eruptions}. The first two types of variations occur with approximately constant bolometric magnitude, but this is not the case for the third case. It is important to emphasize that these values of $\Delta V$ were derived for Galactic LBVs and can be lower for low metallicities objects. A star needs to be observed during an eruption, presenting both photometric and spectral variations, to be classified as LBV, otherwise, it can be only classified as an LBV candidate (LBVc).

However, the physical mechanism that triggers the eruptions is still not well known. It could be related to high rotation \citep{2009ApJ...705L..25G}, radiative pressure with a modified Eddington limit \citep{1988ApJ...324..279L}, turbulent pressure \citep{1984A&A...138..246D}, internal dynamic mechanisms \citep{1993ApJ...408L..85S, 1993MNRAS.263..375G}, binarity, or a combination of some of these mechanisms.

Luminous blue variables are characterized by high mass-loss rates ($10^{-7} \leq \dot{M} \leq 10^{-3}$ M$_{\odot}$\,yr$^{-1}$), low or not very high expansion velocities ($v_{\infty}$ from 50 to a few hundreds of km s$^{-1}$), high luminosities ($\log{L/L_{\odot}} > 5.5$) and high effective temperatures ($8000\,\mathrm{K} \leq T_{\mathrm{eff}} \leq 30000$~K). As noted by \citet{1994PASP..106.1025H}, LBVs are located in two main strips in the HR diagram, related to the quiescent and eruptive stages. The effective temperature in the eruption phase seems to be constant for all LBVs, but during the quiescence it seems to be a function of the stellar luminosity. 

In this paper, we present the detection of new eruptions for two LBVs in the Magellanic Clouds R\,40 and R\,110, based on new spectroscopic and photometric data. In Sect.~\ref{sect:sample}, we describe these two stars. In Sect.~\ref{sec:obs}, we present our observations and the public data used in our work. In Sect.~\ref{sec:analyses}, we analyze the data and discuss the results for each star. The conclusions of the paper are presented in Sect.~\ref{sec:conc}.

\section{Our sample}\label{sect:sample}

In this work, we present the detection of new eruptions in two LBVs from the Magellanic Clouds \object{RMC\,40} (SMC) and \object{RMC\,110} (LMC).

\object{RMC\,40} (\object{LHA 115-S 52}, \object{HD 6884}), hereafter R\,40, was the first LBV detected in SMC \citep{1993A&A...280..508S}. Its $V$ magnitude was reported as 10.73 mag by \citet{1960MNRAS.121..337F} and 10.48 mag by \citet{1985A&AS...61..237S}. \citet{1993A&A...280..508S} reported a $V$ magnitude of 10.25 mag and they showed the spectral change of R\,40 from B8Ie \citep{1960MNRAS.121..337F} to A4 in 1993, characterizing a typical LBV eruption. The visual brightness of R\,40 reached a maximum of 9.8 mag around 1996 \citep{1998A&A...333..565S} and after that it started to decrease. 
The physical parameters determined by \citet{1993A&A...280..508S} are $T_{\mathrm{eff}} = 8700$~K, $\log{g}= 0.75$, $M_{\mathrm{Bol}} = -9.4$, $R = 280\,R_\odot$, and a current mass of $M = 16\,M_\odot$. The identification of this LBV in the SMC indicates that low metallicity environments do not prevent massive stars from evolving to this phase.

\object{RMC\,110} (\object{LHA 120-S 116}, \object{HD 269662}), hereafter R\,110, was identified as LBV by \citet{1990A&A...228..379S}. R\,110 was the faintest LBV classified at that time and was classified as $M_{\mathrm{Bol}} = -8.9$. The first suspicion of its variability was cited by \citet{1984A&A...140..459S}. A discrepancy in the B9Ieq classification from \citet{1960MNRAS.121..337F} was shown by \citet{1985A&AS...61..237S}, in which R\,110 was classified as an early A-type star. According to \citet{1990A&A...228..379S}, the visual magnitude of this star increased from $V=10.5$ mag in December 1980 to 9.99 mag in January 1989 and the  spectral type changed from B to F0. In contrast to the other LBVs, this star has not shown significant mass-loss rate variation between the quiescent and eruptive states; this value is  $\dot{M} = 10^{-6}$~M$_\odot$\,yr$^{-1}$. The physical parameters derived by \citet{1990A&A...228..379S} are $T_{\mathrm{eff}} = 7600$ K, $\log{g}= 0.45$, $R = 310 \,R_\odot$ and a current mass of $M = 10\,M_\odot$.

\section{Observations}\label{sec:obs}

\subsection{High-resolution spectroscopy}

We observed these two stars with the \textit{Fiber-fed Extended Range Optical Spectrograph} (FEROS) mounted at 2.2 m ESO-MPI telecope at La Silla Observatory (Chile). The instrumental configuration provides a resolution of $0.03$\,\AA/pixel (R $\sim$48000) in a spectral range from 3600 to 9200\,\AA. The spectra taken in 2005 were reduced using MIDAS routines developed by our group, following standard echelle reduction procedures. The data taken between 2007 and 2016 were reduced by the ESO/FEROS pipeline. The S/N ratio is between 60 and 120 around H$\alpha$.

We complemented our data with public pipeline reduced spectra from FEROS and the \textit{Ultraviolet and Visual Echelle Spectrograph} (UVES) obtained from the ESO Archive. The UVES spectra have a resolution from $40,000$ to $100,000$ in a spectral range from 3000 to 11000\,\AA, which varies according the observing configuration. Table~\ref{tab:spec_list} lists all the spectra.

\begin{table}
        \begin{center}
                \caption{List of high-resolution spectra used in our study. The spectra marked with an asterisk were observed by us and the others were obtained from ESO public data archive. The exposure time is associated with the number of spectra obtained.}
                \label{tab:spec_list}
                \footnotesize
                \begin{tabular}{cccc}
                        \hline
                                  Star           & Inst. &    Date     & t$_{\mathrm{exp}}$ (s) \\ \hline\hline
                        \multirow{9}{*}{RMC 40}  & UVES  & 2000-07-09  &     $1 \times 600$     \\
                                                 & UVES  & 2002-05-06  &     $1 \times 360$     \\
                                                 & FEROS & 2005-12-12* &     $2 \times 450$     \\
                                                 & FEROS & 2006-10-11  &     $2 \times1000$     \\
                                                 & FEROS & 2007-10-05* &     $2 \times 450$     \\
                                                 & FEROS & 2008-08-04  &     $2 \times1800$     \\
                                                 & FEROS & 2014-11-28* &     $2 \times 700$     \\
                                                 & FEROS & 2015-12-04* &     $2 \times 500$     \\
                                                 & FEROS & 2016-11-02* &     $2 \times 500$     \\ \hline
                        \multirow{4}{*}{RMC 110} & FEROS & 2005-12-12* &     $2 \times 450$     \\
                                                 & FEROS & 2007-02-20  &     $7 \times 200$     \\
                                                 & FEROS & 2014-12-04* &     $2 \times 400$     \\
                                                 & FEROS & 2016-01-13* &     $2 \times 700$     \\ \hline
                \end{tabular}
        \end{center}
\end{table}

Normalization, cosmic ray cleaning, and equivalent width measurements were done using standard IRAF tasks\footnote{IRAF is distributed by the National Optical Astronomy Observatories, which are operated by the Association of Universities for Research in Astronomy, Inc., under cooperative agreement with the National Science Foundation. See \url{http://iraf.noao.edu/}}.


\subsection{Photometry}

We performed 12 observing campaigns from 2014 to 2017 at Observat\'orio do Pico dos Dias (OPD/LNA, Brazil), using the 0.6~m telescope Boller \& Chivens, with 21 nights of effective observation. For these observations we used the CCD Andor IkonL with an E2V CCD42-40 detector (2048$\times$2048 13.5~$\mu$m square pixels). The stars were observed using $UBVRI$ filters. The observations were reduced via our own routines for PSF photometry. To convert the instrumental flux to magnitudes, we compared values of field stars taken from the UCAC4 catalog \citep{2013AJ....145...44Z} for $B$ and $V$ filters, DENIS catalog \citep{2000A&AS..144..235C} for $I$ filter, and stars taken from other catalogs from the Simbad database (\url{http://simbad.u-strasbg.fr/}) for $R$ and $U$ filters, using an iterative Monte Carlo comparision method. There, hundreds of groups of randomly chosen stars in the catalogs were compared to R\,40 and R\,110, obtaining their magnitudes. We assumed the median of these values. The error is estimated as the median absolute deviation from the derived magnitudes. The results of our photometric observations are shown in Tables \ref{tab:phot_R40} and \ref{tab:phot_R110}.

\begin{table*}
        \centering
        \caption{$UBVRI$ photometric observations for our sample obtained in the period of 2014-2017 at OPD/LNA.}
        \subcaption{Observations for R\,40.} 
        \label{tab:phot_R40}
        \begin{tabular}{ccccccc}
                \hline
                Date    &    JD     &      $U$       &      $B$       &      $V$      &      $R$      &      $I$      \\ \hline\hline
                2014-10-23 & 2456953.6 & 9.98$\pm$0.08  & 9.94$\pm$0.02  & 9.34$\pm$0.04 & 9.36$\pm$0.05 & 8.78$\pm$0.02 \\
                2014-10-28 & 2456958.5 & 10.10$\pm$0.04 & 10.08$\pm$0.01 & 9.46$\pm$0.03 & 9.25$\pm$0.04 & 8.89$\pm$0.02 \\
                2014-10-29 & 2456959.5 & 10.07$\pm$0.04 & 9.91$\pm$0.03  & 9.36$\pm$0.03 & 9.21$\pm$0.05 & 8.84$\pm$0.02 \\
                2014-11-11 & 2456972.5 &      ---       &      ---       & 9.37$\pm$0.04 & 9.28$\pm$0.04 &      ---      \\
                2014-11-12 & 2456973.5 &      ---       &      ---       & 9.39$\pm$0.04 &      ---      & 8.86$\pm$0.02 \\
                2014-11-17 & 2456978.5 &      ---       &      ---       & 9.42$\pm$0.03 &      ---      &      ---      \\
                2015-10-05 & 2457300.6 &      ---       & 9.95$\pm$0.02  & 9.30$\pm$0.03 & 9.11$\pm$0.05 &      ---      \\
                2015-10-06 & 2457301.5 & 10.08$\pm$0.06 & 9.95$\pm$0.02  &      ---      &      ---      &      ---      \\
                2015-10-07 & 2457302.5 & 10.08$\pm$0.05 & 9.95$\pm$0.03  & 9.32$\pm$0.03 & 9.27$\pm$0.04 & 8.73$\pm$0.02 \\
                2015-10-08 & 2457303.5 & 10.06$\pm$0.08 & 9.93$\pm$0.03  & 9.31$\pm$0.03 & 9.19$\pm$0.05 & 8.71$\pm$0.01 \\
                2015-10-08 & 2457304.4 &      ---       & 9.93$\pm$0.02  & 9.36$\pm$0.03 & 9.22$\pm$0.05 & 8.74$\pm$0.01 \\
                2015-12-12 & 2457368.5 &      ---       & 9.92$\pm$0.03  & 9.30$\pm$0.03 & 9.22$\pm$0.05 &      ---      \\
                2016-01-25 & 2457413.4 & 10.02$\pm$0.08 & 9.87$\pm$0.03  & 9.27$\pm$0.03 & 9.09$\pm$0.05 & 8.66$\pm$0.02 \\
                2016-07-26 & 2457595.8 & 10.04$\pm$0.04 & 9.90$\pm$0.03  & 9.18$\pm$0.03 & 9.15$\pm$0.05 & 8.61$\pm$0.02 \\
                2016-07-27 & 2457596.8 &      ---       & 9.89$\pm$0.03  & 9.20$\pm$0.04 & 8.98$\pm$0.05 & 8.62$\pm$0.02 \\
                2016-10-29 & 2457691.4 &      ---       &      ---       & 9.16$\pm$0.02 & 8.97$\pm$0.06 &      ---      \\
                2016-12-17 & 2457739.5 &      ---       &      ---       & 9.40$\pm$0.02 &      ---      &      ---      \\
                2016-12-18 & 2457740.6 &      ---       & 10.08$\pm$0.03 & 9.37$\pm$0.03 &      ---      &      ---      \\
                2016-12-22 & 2457744.6 &      ---       & 10.06$\pm$0.03 & 9.30$\pm$0.03 &      ---      &      ---      \\
                2017-02-23 & 2457807.5 &      ---       & 10.21$\pm$0.03 & 9.34$\pm$0.03 & 9.19$\pm$0.04 & 8.63$\pm$0.02 \\ \hline
        \end{tabular}
        
        \bigskip
        
        \subcaption{Observations for R\,110.}
        \label{tab:phot_R110}
        \begin{tabular}{ccccccc}
                \hline
                Date    &    JD     &      $U$       &      $B$       &      $V$       &      $R$       &      $I$       \\ \hline\hline
                2014-10-23 & 2456953.7 & 10.43$\pm$0.08 & 10.70$\pm$0.04 & 10.45$\pm$0.05 & 10.20$\pm$0.06 & 10.26$\pm$0.02 \\
                2014-10-29 & 2456959.8 &      ---       & 10.68$\pm$0.03 & 10.46$\pm$0.04 & 10.18$\pm$0.06 & 10.24$\pm$0.02 \\
                2014-11-11 & 2456972.7 &      ---       &      ---       & 10.45$\pm$0.04 &      ---       &      ---       \\
                2014-11-12 & 2456973.6 & 10.11$\pm$0.04 & 10.64$\pm$0.04 & 10.43$\pm$0.05 & 10.23$\pm$0.02 & 10.18$\pm$0.02 \\
                2014-11-17 & 2456978.8 &      ---       &      ---       & 10.41$\pm$0.05 &      ---       &      ---       \\
                2015-10-05 & 2457300.7 &      ---       & 10.70$\pm$0.03 & 10.47$\pm$0.04 & 10.23$\pm$0.07 & 10.25$\pm$0.01 \\
                2015-10-07 & 2457302.6 & 10.10$\pm$0.04 &      ---       & 10.48$\pm$0.03 & 10.28$\pm$0.06 & 10.22$\pm$0.01 \\
                2016-12-18 & 2457740.7 &      ---       & 10.80$\pm$0.04 & 10.57$\pm$0.05 &      ---       &      ---       \\
                2016-12-23 & 2457745.7 &      ---       &      ---       & 10.53$\pm$0.04 &      ---       &      ---       \\
                2017-02-22 & 2457806.5 &      ---       & 10.80$\pm$0.04 & 10.55$\pm$0.04 &      ---       & 10.38$\pm$0.03 \\ \hline
        \end{tabular} 
\end{table*}

Photometric data obtained from the literature and public archives were also used to obtain light curves (LC) for these two objects. We collected data from various photometric sources: \textit{American Association of Variable Star Observers}\footnote{For a better analysis of the light curve of R\,110, we also included the data from AAVSO Visual band, which is similar to the $V$ filter, but is not precise and has to be analyzed carefully.} (AAVSO); \textit{All Sky Automated Survey} (ASAS); \textit{Long-Term Photometry Variables at ESO} \citep[LTPV,][]{1991A&AS...87..481M, 1993A&AS..102...79S, 1995A&AS..109..329M, 1995A&AS..113...31S} and from \citeauthor{1998JAD.....4...10V} (\citeyear{1982A&AS...47..591V, 1998JAD.....4...10V}). Other photometric data taken from the literature are described in Table~\ref{tab:phot_lit}.


\begin{table*}
        \centering
        \caption{$UBVRI$ photometry for R~40 and R~110 obtained from the literature.}
        \label{tab:phot_lit}
        \begin{tabular}{cccccccc}
                \hline
                Star & Reference & Date & $V$ & $B-V$ & $U-B$ & $V-R$ & $V-I$ \\
                \hline\hline
                \multirow{8}{*}{RMC 40} & {\citet{1960MNRAS.121..337F}} & 1955-1960       & 10.73 & 0.08 & ---   & ---  & ---  \\
                & {\citet{1970A&A.....9...95D}} & Sep.-Dec., 1968 & 10.58 & 0.07 & -0.58 & ---  & ---  \\
                & {\citet{1972DunOP...1..133B}} & 1966            & 10.61 & 0.10 & -0.56 & ---  & ---  \\
                & {\citet{1973ApJ...181..327O}} & 1970-1971       & 10.51 & ---  & ---   & ---  & ---  \\
                & {\citet{1977A&AS...30..261A}} & 1974            & 10.52 & 0.10 & -0.57 & ---  & ---  \\
                & {\citet{1980A&AS...42....1A}} & 1971-1978       & 10.49 & 0.12 & -0.61 & ---  & ---  \\
                & {\citet{1985A&AS...61..237S}} & Aug. 1983       & 10.48 & 0.12 & -0.59 & ---  & ---  \\
                & {\citet{1993A&A...280..508S}} & Nov. 9, 1991    & 10.25 & 0.16 & -0.45 & ---  & ---  \\\hline
                \multirow{9}{*}{RMC 110}& {\citet{1970BOTT....5..269M}} & 1969            & 10.41 & 0.24 & ---   & 0.32 & 0.50 \\
                & {\citet{1977A&AS...30..245W}} & 1969-1975       & 10.57 & 0.19 & ---   & ---  & ---  \\
                & {\citet{1990A&A...228..379S}} & Dec. 1980       & 10.50 & ---  & ---   & ---  & ---  \\
                & {\citet{1990A&A...228..379S}} & Jan. 1982       & 10.27 & 0.24 & -0.27 & ---  & ---  \\
                & {\citet{1990A&A...228..379S}} & Aug. 1983       & 10.40 & 0.23 & -0.32 & ---  & ---  \\
                & {\citet{1990A&A...228..379S}} & Jan. 1984       & 10.34 & 0.23 & -0.35 & ---  & ---  \\
                & {\citet{1990A&A...228..379S}} & Aug. 1984       & 10.31 & 0.25 & -0.29 & ---  & ---  \\
                & {\citet{1990A&A...228..379S}} & Dec. 1984       & 10.20 & 0.26 & -0.27 & 0.19 & 0.40 \\
                & {\citet{1990A&A...228..379S}} & Jan. 1989       &  9.99 & 0.35 & -0.08 & 0.22 & 0.46 \\
                \hline
        \end{tabular}
\end{table*}

These data are available in the Johnson photometric system, except the data from the LTPV survey, which uses standard Str\"omgren filters. Thus, we converted the LTPV data to the Johnson system, via the relations provided by \citet{2001A&A...369.1140H}.

\section{Data analysis}\label{sec:analyses}

\subsection{R\,40}

\subsubsection{Light curve}

\begin{figure*}
        \centering
        \includegraphics[width=\linewidth]{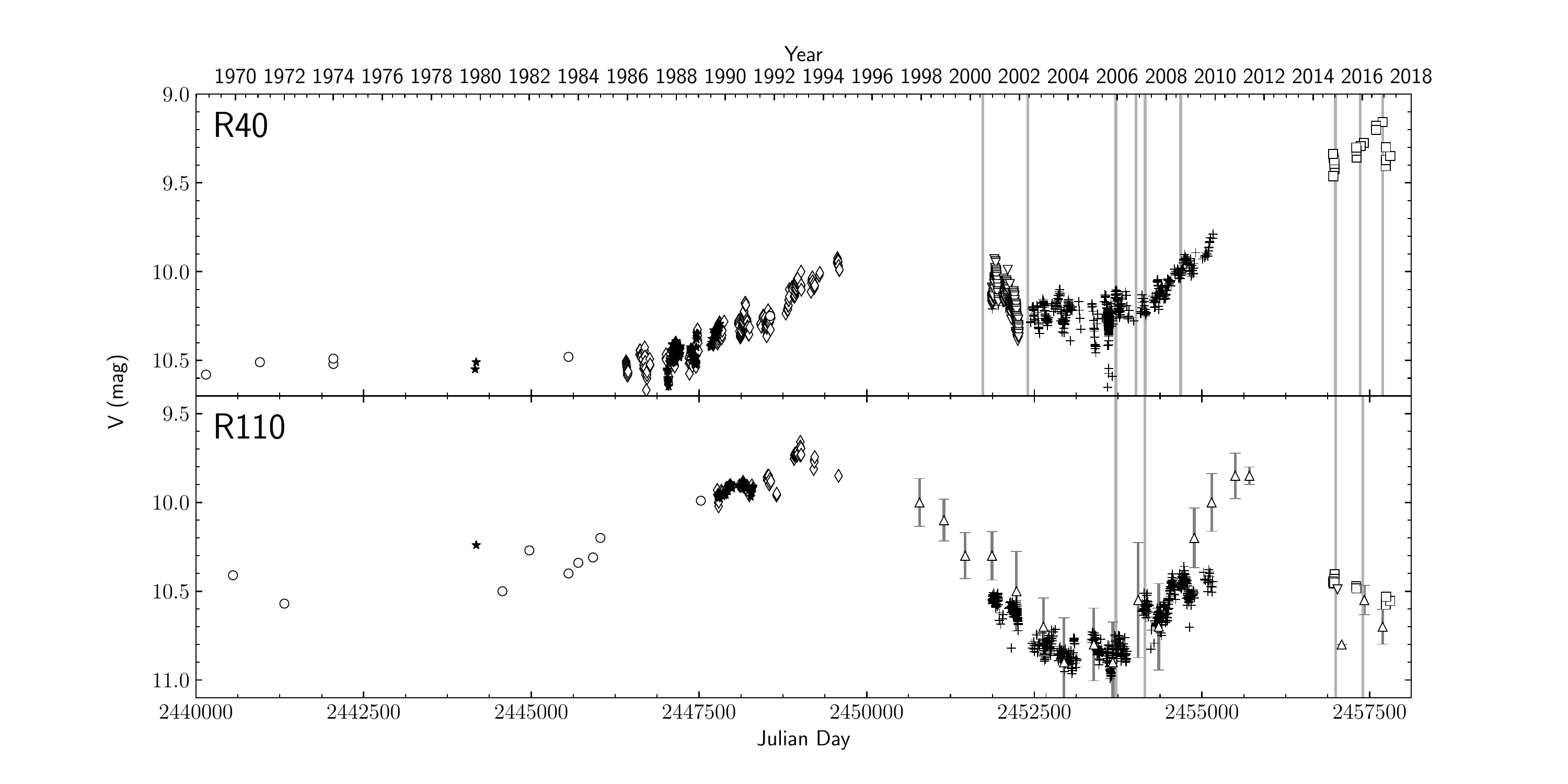}
        \caption{Light curves of R\,40 and R\,110. Squares indicate OPD/LNA (this work); diamonds indicate LTPV; crosses indicate ASAS; downward triangles indicate AAVSO ($V$); stars indicate \citeauthor{1998JAD.....4...10V}; circles indicate references cited in Table~\ref{tab:phot_lit}; and upward triangles indicate the median of the AAVSO (Vis.) data of each year with standard deviation error bars. The vertical lines represent the observation dates of the spectra listed in Table~\ref{tab:spec_list}.}
        \label{fig:light_curves}
\end{figure*}

In Fig.~\ref{fig:light_curves}, we show the light curve for R\,40. The first reported eruption for R\,40, which was identified by \citet{1993A&A...280..508S}, is clearly visible. This eruption started in approximately 1987, probably reaching the maximum in 1996 and had a $V$ magnitude of approximately $9.8$ \citep{1998A&A...333..565S}. However, there is a lack of data from 1996 until 2000, hampering the confirmation of this maximum estimate. We identified a new eruption that started in 2005, as can also be seen in Fig.~\ref{fig:light_curves}; this eruption has not previously been reported in the literature. This is an ongoing eruption, where the maximum brightness was probably reached in the middle of 2016, when R\,40 had $V \sim 9.2$~mag. \citet{1993A&A...280..508S} mentioned that R\,40 was the brightest star in the SMC during the first eruption, and since it is even brighter now, it most likely continues to be the most luminous star in that galaxy. 

After this probable maximum, the light curve has a drop off almost $0.1$~mag just few months later. This fast decrease is not expected to happen, especially compared to the behavior of other eruptions, and its cause is not known.

As we can also see in Fig.~\ref{fig:light_curves}, the minimum brightness just before each one of these two eruptions has a different value. Previous to the 1987 eruption, R\,40 appears to have had a constant magnitude in the $V$ band, $V\sim10.6$, in contrast to $V\sim10.2$ seen in 2004-2006 just before the new eruption. This may indicate some contribution from the material ejected during the first eruption. 

\subsubsection{Spectral characteristics}

\begin{figure*}
        \centering
        \includegraphics[width=\linewidth]{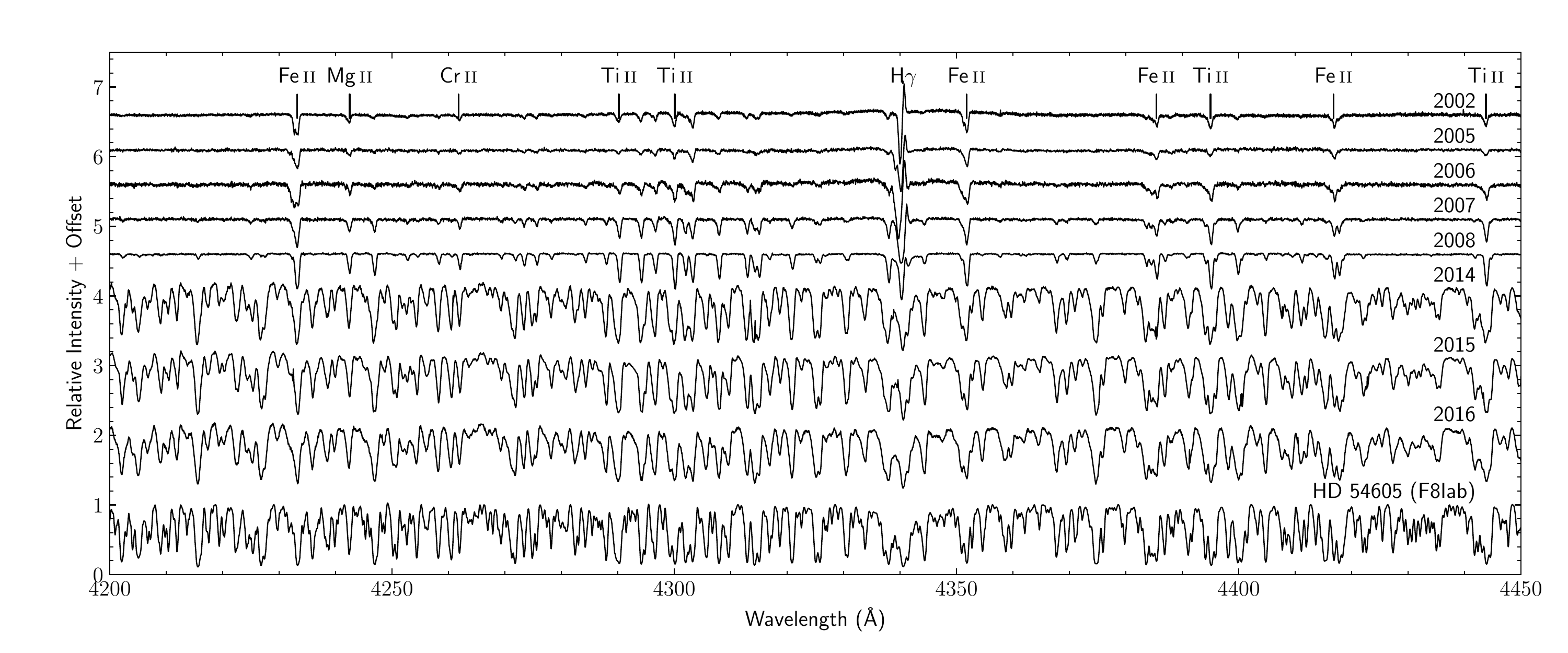}
        \caption{Spectral variation of R\,40 seen in the spectra taken from 2002 (top) until 2016 (bottom). We can see the change from a late B/early A-type to a late F-type spectrum, implying a new eruption. The spectrum of HD~54605 (F8Ia) is also shown for comparison.}
        \label{fig:R40_spec_change}
\end{figure*}

In total, we analyzed nine high-resolution spectra (seven FEROS and two UVES) to study the spectral variations of R\,40 during the last 16 years. We identified two different behaviors: one for the spectra taken from 2000 until 2008, and another for those spectra taken in 2014-2016 (see Fig.~\ref{fig:R40_spec_change} and ~\ref{fig:A1}).

\subsubsection*{2000-2008 (quiescence):}

The spectra taken from 2000 until 2008, which correspond to the period after the first eruption and just before the new eruption, the so-called {\it quiescence}, do not have significant variations. They are similar to the spectrum of a B-A supergiant (see Fig.~\ref{fig:A1}), as previously reported by \citet{1993A&A...280..508S}. 

From the spectrum taken in 2005, we also derived a mean radial velocity of $169 \pm 10$~km\,s$^{-1}$, obtained from lines of singly ionized metals in pure absorption, especially \ion{Fe}{ii}, \ion{Cr}{ii}, \ion{Ti}{ii,} and \ion{Mg}{ii}. This value is lower than that derived by \citet{1993A&A...280..508S}, $189 \pm 5$~km\,s$^{-1}$, obtained using lines of the same ions. This fact means that the lines observed by \citet{1993A&A...280..508S} may have been formed in an expanding shell and not in the photosphere of the star. \citet{1960MNRAS.121..337F} also reported a radial velocity of $181$~km\,s$^{-1}$ obtained from spectra taken before 1960.

In the spectra taken in 2006-2008 (see Fig.~\ref{fig:R40_spec_change} and ~\ref{fig:A1}), the lines of neutral and singly ionized metals clearly became more intense, indicating a gradual decrease in temperature and the beginning of the new ongoing eruption. Absorption lines of singly ionized metals, such as \ion{Fe}{ii}, \ion{Cr}{ii}, \ion{Ti}{ii,} and \ion{Si}{ii}, dominate the blue part of the spectrum. 

A few \ion{Fe}{ii} lines have P~Cygni profiles in the spectra taken in 2005-2008, but not earlier. These profiles present one or two variable absorption components. The same is seen for Balmer lines, which present intense P~Cygni profiles (Fig~\ref{fig:S52_Halpha}) with variabilities in both emission and absorption components.

Forbidden lines are commonly seen in the quiescent phase of LBVs, but they are not seen in our R\,40 spectra. This is probably caused by a remaining effect of the first eruption, causing an increase of density and avoiding the formation of these lines. However, \citet{1985A&AS...61..237S} also reported the absence of forbidden lines during the previous quiescent stage, i.e., before the first identified eruption.

\subsubsection*{2014-2016 (ongoing eruption):}

The spectra taken from 2014 until 2016 changed from the previous B or A type to a late F-type, corresponding to the probable maximum of the new ongoing eruption and the formation of a pseudo-photosphere. From the visual comparison with spectra from the ESO UVES-POP catalog \citep{2003Msngr.114...10B}, we noted that they are very similar to the F8Iab type spectrum of \object{HD~54605}\footnote{This star, also named \textit{Wezen}, $\delta$~CMa, or HR~2693 ($V=1.84$~mag, $T_{\mathrm{eff}} = 6476$~K, $\log{g} = 0.89,$ and [Fe/H]$=+0.28$, see \citet{2014AJ....147..137L}), is a bright galactic yellow supergiant.} (Fig.~\ref{fig:R40_spec_change} and ~\ref{fig:A1}). This late F-type spectrum is very uncommon during an LBV eruption, which is normally associated with a late A or early F-type.

Concerning H$\alpha$ and H$\beta$, they both show weak P~Cygni profiles with some variability (see Fig~\ref{fig:S52_Halpha}). On the other hand, the other Balmer lines are seen in pure absorption (Fig~\ref{fig:A1}). Regarding \ion{He}{i} lines, in contrast to what was reported by \citet{1993A&A...280..508S} during the first eruption, these lines have completely disappeared in the spectra taken in 2014-2016.

Another dramatic change is related to \ion{Ca}{ii} H and K lines (Fig.~\ref{fig:A1}). Until 2008, during the quiescent phase, these lines were narrow and had a clear separation of the interstellar and circumstellar components. This shape completely changes in the 2014-2016 spectra, in which they appear broad and intense and with the different (stellar plus interstellar) components blended, very similar to a F8Iab star spectrum. The low excitation potential of these lines indicates their formation within a cool pseudo-photosphere, as seen during the eruption of other LBVs \citep{2013A&A...555A.116M}.

In addition, in the red part of the spectra ($>7000$\,\AA), the Paschen lines did not have strong variations in the quiescent and eruptive stages, appearing as single-absorption profiles. Some lines from \ion{Ca}{i}, \ion{Mn}{ii}, \ion{Fe}{i}, \ion{Fe}{ii}, \ion{S}{i}, \ion{S}{ii}, \ion{Ti}{i}, \ion{Ti}{ii}, and some s- and r-process elements, such as \ion{La}{ii}, \ion{Eu}{ii,} and \ion{Ba}{ii} (and possibly \ion{Sr}{ii}, \ion{Sc}{ii}, \ion{Nd}{ii}, \ion{Y}{i}, \ion{Gd}{ii}, \ion{V}{i}, and \ion{Yb}{i}), appear only in the 2014-2016 spectra (see Fig.~\ref{fig:A1}). 

A radial velocity of $172.5$\,km~s$^{-1}$ was derived using the center of the metal lines present in the spectrum taken in 2016.

\subsubsection{Physical parameters of R\,40}

Deriving the precise spectral type for B-A stars, such as LBVs in the quiescent phase, is not an easy task. Using some empirical spectroscopic criteria, based on equivalent width ratios, we can estimate the spectral classification for B supergiants in the Magellanic Clouds \citep{1997A&A...317..871L}. These criteria were later extended to G-type supergiants by \citet{2003MNRAS.345.1223E} and \citet{2004MNRAS.353..601E}. However for LBVs, the lines can be largely affected by wind contribution, making this classification very uncertain, especially considering Balmer and \ion{Fe}{ii} lines, which are mainly formed in the wind. Thus, we preferred to use the calibration based on \ion{Mg}{ii}~$4482\,\AA$/\ion{He}{i}~$4471\,\AA$ equivalent widths ratio from \citet{1997A&A...317..871L}, \citet{2003MNRAS.345.1223E}, and \citet[Fig.~3]{2008A&A...487..697K}, which are probably less affected by the wind.  


\begin{table*}
\centering
\caption{Effective temperature estimations for R\,40 and R\,110, based on different spectroscopic and photometric criteria, where >A2 corresponds to spectral types later than A2.}
\label{tab:R40_colortemperature}
\begin{tabular}{cccccccccc}
        \hline
                                                                                             \multicolumn{9}{c}{R\,40}                                                                                       &  \\ \hline\hline
          Year   & \ion{Mg}{ii}~$4482 \AA$/\ion{He}{i}~$4471 \AA$ & Spectral &  Spectral  &  $(B-V)_0$  & $T_\mathrm{eff}$~(K)$^1$ & \multicolumn{2}{c}{$T_\mathrm{eff}$~(K)$^2$} & $T_\mathrm{eff}$~(K)$^3$ &  \\
                 &                                                & type$^1$ &    type    &             &                          & HM84  &                 EH03                 &                          &  \\ \hline\hline
          1960   &                                                &          & B8Ieq$^4$  &             &                          & 10900 &                12000                 &                          &  \\
          1960   &                                                &          &            & $-$0.03$^4$ &                          &       &                                      &          11100           &  \\
          1966   &                                                &          &            & $-$0.01$^5$ &                          &       &                                      &          10500           &  \\
          1968   &                                                &          &            & $-$0.04$^6$ &                          &       &                                      &          11400           &  \\
          1974   &                                                &          &            & $-$0.01$^7$ &                          &       &                                      &          10500           &  \\
          1983   &                                                &          &            &  0.01$^8$   &                          &       &                                      &          10000           &  \\
          1991   &                                                &          &            &  0.05$^9$   &                          &       &                                      &           9200           &  \\
          1994   &                                                &          &   A4$^9$   &             &                          & 8800  &                 8000                 &                          &  \\
          2002   &                4.70 $\pm$ 1.14                 &    A2    &            &             &           8500           &       &                                      &                          &  \\
          2005   &                3.77 $\pm$ 0.73                 &  A0-A2   &            &             &           9000           &       &                                      &                          &  \\
          2005   &                                                &          & A0$\sim$A2 &             &                          & 9300  &                 9000                 &                          &  \\
          2006   &                5.40 $\pm$ 1.38                 & $\sim$A2 &            &             &           8500           &       &                                      &                          &  \\
          2007   &                7.00 $\pm$ 2.29                 &   >A2    &            &             &          <8500           &       &                                      &                          &  \\
          2008   &                6.50 $\pm$ 0.94                 &   >A2    &            &             &          <8500           &       &                                      &                          &  \\
          2014   &                                                &          &            &    0.48     &                          &       &                                      &           6200           &  \\
          2015   &                                                &          &            &    0.51     &                          &       &                                      &           6100           &  \\
          2016   &                                                &          &   F8Iab    &             &                          & 6200  &                 5750                 &                          &  \\
        Jun/2016 &                                                &          &            &    0.57     &                          &       &                                      &           5900           &  \\
        Dec/2016 &                                                &          &            &    0.65     &                          &       &                                      &           5800           &  \\ \hline\hline
                                                                                             \multicolumn{9}{c}{R\,110}                                                                                      &  \\ \hline\hline
          1960   &                                                &          & B9Ieq$^4$  &             &                          & 10250 &                10500                 &                          &  \\
          1969   &                                                &          &            & 0.04$^{10}$ &                          &       &                                      &           9400           &  \\
          1982   &                                                &          &            &  0.04$^8$   &                          &       &                                      &           9400           &  \\
          1984   &                                                &          &            &  0.06$^8$   &                          &       &                                      &           9000           &  \\
          1989   &                                                &          &            &  0.15$^8$   &                          &       &                                      &           7800           &  \\
          1989   &                                                &          &  F0Ia$^8$  &             &                          & 7800  &                 6750                 &                          &  \\
          2005   &                                                &          &     B9     &             &                          & 10250 &                10500                 &                          &  \\
          2005   &                1.95 $\pm$ 0.34                 &    B9    &            &             &          10500           &       &                                      &                          &  \\
          2007   &                2.69 $\pm$ 0.63                 &  A0-A2   &            &             &           9000           &       &                                      &                          &  \\
          2014   &                3.54 $\pm$ 0.89                 &  A0-A2   &            &             &           9000           &       &                                      &                          &  \\
          2014   &                                                &          &   A0-A2    &             &                          & 9300  &                 9000                 &                          &  \\
          2014   &                                                &          &            & 0.03$^{11}$ &                          &       &                                      &           9600           &  \\
          2016   &                5.00 $\pm$ 1.70                 &    A2    &            &             &           8500           &       &                                      &                          &  \\
          2016   &                                                &          &     A2     &             &                          & 9100  &                 8500                 &                          &  \\
          2016   &                                                &          &            &    0.03     &                          &       &                                      &           9600           &  \\ \hline
\end{tabular} 
\tablefoot{ 1-Spectral type and $T_\mathrm{eff}$ were estimated from \ion{Mg}{ii}~$4482 \AA$/\ion{He}{i}~$4471 \AA$ ratio; these measurements were only obtained for spectra for which the lines are not blended; 2-$T_\mathrm{eff}$ was estimated using the relation for spectral types from \citet[Tab.~2]{1984ApJ...284..565H} (HM84) and \citet[Tab. 5]{2003MNRAS.345.1223E} (EH03); and 3-$T_\mathrm{eff}$ was estimated from $(B-V)_0$ . Owing to the uncertainty in the reddening, the error is $\sigma_{T_{\mathrm{eff}}} = 1000$~K for $T_{\mathrm{eff}} \geq 9000$~K and $\sigma_{T_{\mathrm{eff}}} = 500$~K for $T_{\mathrm{eff}} < 9000$~K). The references are 4-\citet{1960MNRAS.121..337F}, 5-\citet{1972DunOP...1..133B}, 6-\citet{1970A&A.....9...95D}, 7-\citet{1977A&AS...30..261A}, 8-\citet{1985A&AS...61..237S}, 9-\citet{1993A&A...280..508S}, 10-\citet{1970BOTT....5..269M}, and 11-AAVSO. The other values are from this work. }
\end{table*}

For A-type stars in the SMC specifically, the criteria of \citet{2003MNRAS.345.1223E}, considering the equivalent widths ratio of \ion{Ca}{ii} K/(H$\epsilon$ + \ion{Ca}{ii} H), can also be applied. However, these lines are dependent on $\log{g}$, which may differ in LBVs and normal supergiants, and H$\epsilon$ is affected by the wind. In addition, for all F-type stars with low $\log{g}$, the \ion{Ca}{ii} H and K lines become strong and their ratio goes to 1.

Based on the polynomial relation between intrinsic $(B-V)_0$ color and $T_{\mathrm{eff}}$ proposed by \citet[Table 5]{1996ApJ...469..355F}\footnote{There is a mistake in the caption of that table. In the original caption, the cited polynomial equation is $(B-V)_0 = a + b\log{T_{\mathrm{eff}}} + c\log{T_{\mathrm{eff}}}^2 + ...$, but, as we could not reproduce the results using it, we inverted it to $\log{T_{\mathrm{eff}}} = a + b(B-V)_0 + c(B-V)_0^2 + ...$, thereby reproducing the results correctly.}, we can also estimate $T_{\mathrm{eff}}$ for R\,40, using $B-V$ colors from our various observations (the median for each year) and from the literature. In addition, using the relations between the $T_{\mathrm{eff}}$ and the spectral types from \citet[Tab. 2]{1984ApJ...284..565H} and \citet[Tab. 5]{2003MNRAS.345.1223E}, we could also estimate $T_{\mathrm{eff}}$ of this star in different dates.

The results from these spectroscopic and photometric criteria indicate a strong temperature variability for R 40, from approximately 12000 K in the 1960s, during the quiescence, to 5750 K in 2016 (see Table~\ref{tab:R40_colortemperature}).

We also used the spectroscopic calibration proposed by \citet{2000A&A...358..587K} for F-G supergiants, based on line-depth ratios of some unblended metal lines (see Table 1 in this reference, and the relations number 6, 7 and 22). We obtained $T_{\mathrm{eff}} = 6100 \pm 200$ K for the spectrum taken in 2014, which is similar to the results from the photometric calibration.

Some additional computations were also made using \citet{2010ApJ...719.1784N} polynomial equations, which resulted in an underestimated $T_{\mathrm{eff}}$. This conclusion was also obtained by \citet{2013A&A...555A.116M} for the LMC LBV RMC\,71 (R\,71). Thus, we discarded these results in our analysis.

Concerning the reddening of R\,40, it is not well determined  having been derived by different authors, as $E(B-V)=0.07\sim0.14$~mag \citep{1993A&A...280..508S} and $E(B-V)=0.09\sim0.16$~mag \citep{2011ApJ...737..103S}\footnote{Values taken from \url{http://irsa.ipac.caltech.edu/applications/DUST/}}. For our work, we assume the mean value from both authors of $E(B-V)=0.11\pm0.05$.

\subsubsection*{Modeling using {\sc moog}}

We also decided to perform a spectroscopic analysis of R\,40 using the LTE code {\sc moog} \citep[recent version]{1973PhDT.......180S} and the spectra taken in 2005 and 2014 owing to their high S/N. We chose unblended \ion{Fe}{i} lines to derive $T_{\mathrm{eff}}$, microturbulent velocity and iron abundance. Table~\ref{tab:R40_FeI} shows the \ion{Fe}{i} lines employed in our analysis, their excitation potentials ($\chi$), values of oscillator strengths ($\log{gf}$), and the measured equivalent widths from the spectra taken in 2014. The $\chi$ and $\log gf$ values were taken from \citet{1996ApJS..103..183L} and \citet{1997AJ....114..376C}. Only the lines with equivalent widths smaller than 180~m\AA\ were used for the determination of the atmospheric parameters.

\begin{table}
        \centering
        \caption{\ion{Fe}{i} lines in the spectrum of R\,40 observed in 2014, used to derive $T_{\mathrm{eff}}$ and elemental abundances using the {\sc moog} code. Values for $\chi$ and $\log gf$ were obtained from \citet{1996ApJS..103..183L} and \citet{1997AJ....114..376C}.}
        \label{tab:R40_FeI}
        \begin{tabular}{cccc}
                \hline
                Wavelength (\AA) & $\chi$ (eV) & $\log gf$ & Eq. Width (m\AA) \\ \hline\hline
                5125.12      &    4.22     &  $-$0.08  &        98        \\
                5202.34      &    2.18     &  $-$1.84  &       146        \\
                5281.79      &    3.04     &  $-$0.83  &       120        \\
                5364.87      &    4.45     &   0.23    &       162        \\
                5367.47      &    4.42     &   0.44    &       176        \\
                5373.71      &    4.47     &  $-$0.71  &        23        \\
                5389.48      &    4.42     &  $-$0.25  &        61        \\
                5393.17      &    3.24     &  $-$0.72  &       161        \\
                5400.50      &    4.37     &  $-$0.10  &        83        \\
                5445.04      &    4.39     &   0.04    &       128        \\
                5569.62      &    3.42     &  $-$0.49  &       163        \\
                5576.09      &    3.43     &  $-$0.85  &       109        \\
                5633.95      &    4.99     &  $-$0.12  &        38        \\
                5686.53      &    4.55     &  $-$0.45  &        25        \\
                5717.83      &    4.28     &  $-$0.98  &        43        \\
                5762.99      &    4.21     &  $-$0.41  &        67        \\
                5934.65      &    3.93     &  $-$1.02  &        30        \\
                6020.17      &    4.61     &  $-$0.21  &        31        \\
                6024.06      &    4.55     &  $-$0.06  &        65        \\
                6027.05      &    4.08     &  $-$1.09  &        13        \\
                6056.01      &    4.73     &  $-$0.40  &        34        \\
                6065.48      &    2.61     &  $-$1.53  &        75        \\
                6219.28      &    2.20     &  $-$2.43  &        73        \\
                6230.72      &    2.56     &  $-$1.28  &       143        \\
                6335.33      &    2.20     &  $-$2.18  &        41        \\
                6336.82      &    3.69     &  $-$1.05  &        30        \\
                6344.15      &    2.43     &  $-$2.92  &        21        \\
                6380.74      &    4.19     &  $-$1.32  &        11        \\
                6419.95      &    4.73     &  $-$0.09  &        32        \\
                6569.22      &    4.73     &  $-$0.42  &        23        \\
                6592.91      &    2.72     &  $-$1.47  &        72        \\
                6855.16      &    4.56     &  $-$0.74  &        34        \\ \hline
        \end{tabular}
\end{table}

\begin{figure}
        \centering
        \includegraphics[width=\linewidth]{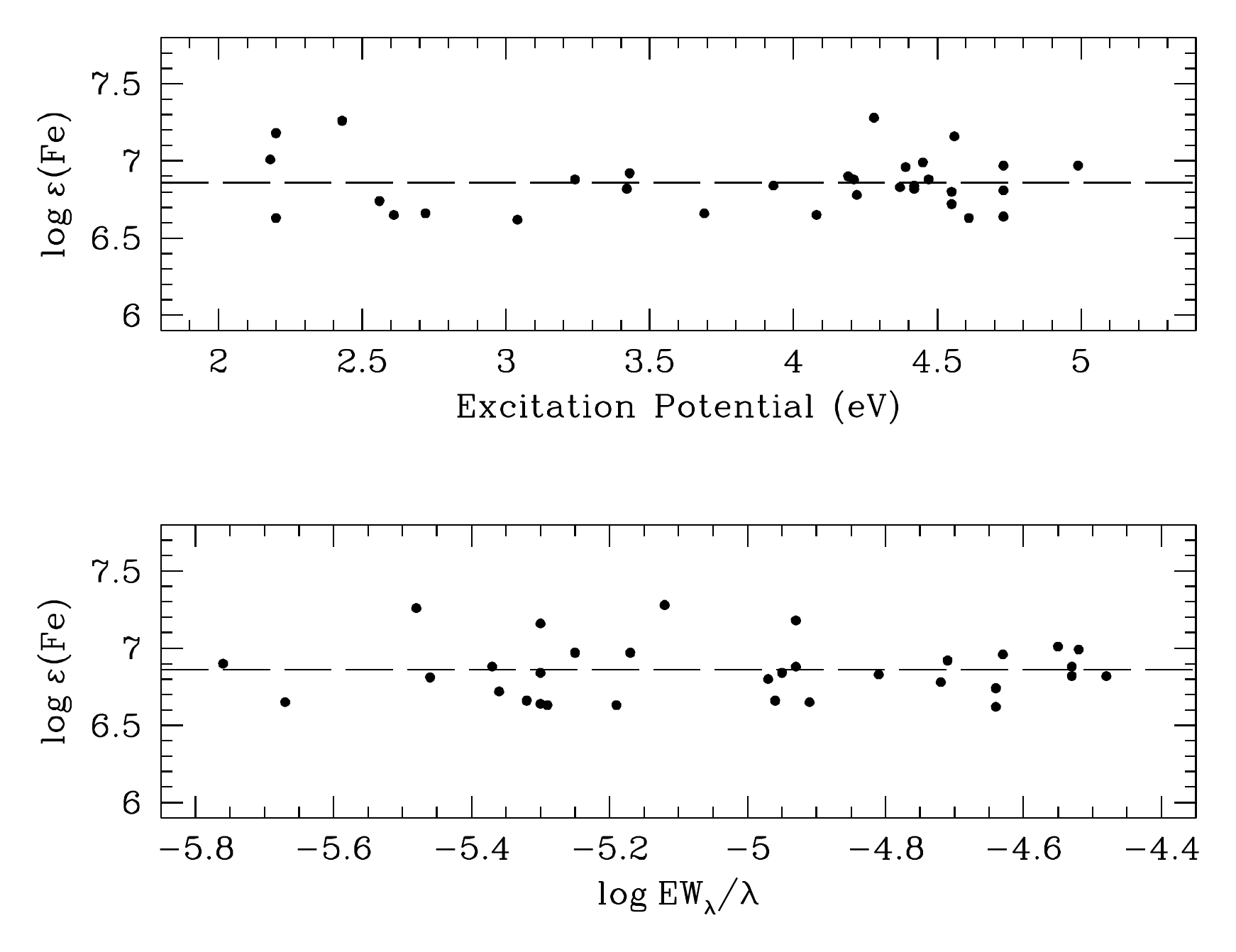}
        \caption{Iron abundance derived from individual \ion{Fe}{i} lines, $\log\varepsilon(\mathrm{Fe})$, using the spectrum of R\,40 taken in 2014, vs. the excitation potential (upper panel) and vs. the reduced equivalent width $\log$(EW$_\lambda/\lambda)$ (lower panel).}
        \label{fig:R40_Fei}
\end{figure}

Assuming $\log g = 0.5$, we estimated $T_{\mathrm{eff}}$ by the solution of the excitation equilibrium, where $T_{\rm eff}$ is derived by the zero slope of the trend between the iron abundances calculated from individual \ion{Fe}{i} lines and their excitation potentials. The microturbulence velocity ($\xi$) was derived by forcing the abundance determined from individual iron lines to be independent of the equivalent width (see Fig.~\ref{fig:R40_Fei}). Using the local thermodynamic equilibrium (LTE) models of \citet{2003IAUS..210P.A20C}, the following atmospheric parameters were derived: $T_{\mathrm{eff}}=6500$~K, $\xi=8.2$~km\,s$^{-1}$, and $\log\varepsilon{\rm (Fe)}=6.86$ corresponding to [Fe/H]=$-$0.64\footnote{We used the notation [X/H]=$\log(N_{\rm X}/N_{\rm H})_\ast - \log(N_{\rm X}/N_{\rm H})_\odot$}, assuming the solar abundance of $\log\varepsilon{\rm (Fe)}_\odot = 7.50$ \citep{2009ARA&A..47..481A}.

The effective temperature derived by this method is in agreement with our previous estimates, but it needs to be seen with caution because strong non-LTE effects and the absence of hydrostatic equilibrium may affect the pseudo-photosphere of R\,40, which may have an even lower $\log g$ than those available by \citet{2003IAUS..210P.A20C} models. In addition, the lines of \ion{Fe}{ii} are very strong and cannot be fitted with the same atmospheric model derived by neutral iron lines. Concerning our metallicity, it is in agreement with results from the literature for the SMC, for example, [Fe/H]=$-0.65\pm0.2$ from \citet{1989ApJS...70..865R}, [Fe/H]=$-$0.68 from \citet{1998AJ....115..605L}, and [Fe/H]=$-$0.73 from \citet{1999ApJ...518..405V}. 

\begin{figure*}
        \centering
        \includegraphics[width=\linewidth]{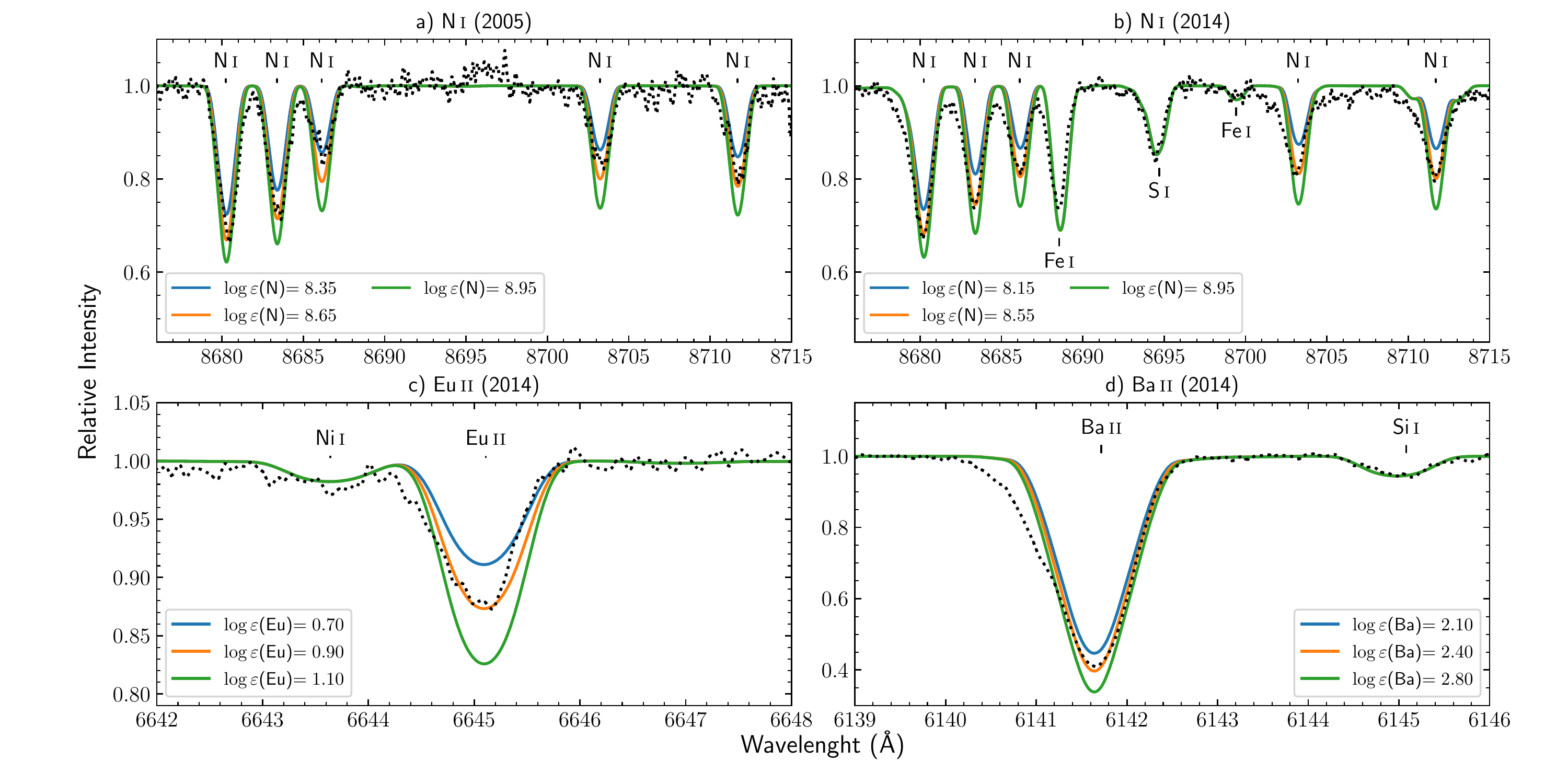}
        \caption{Observed (dotted line) and synthetic (solid lines) spectra of R\,40 in the wavelength regions containing the \ion{N}{i} (a and b), \ion{Eu}{ii} (c), and \ion{Ba}{ii} (d) lines. The observed spectra were obtained in 2005 (a) and 2014 (b, c, and d).}
        \label{fig:R40_MOOG_Abundances}
\end{figure*}

Using {\sc moog}, we also derived the CNO abundances from the observed spectra taken in both the quiescent (2005) and eruptive phases (2014). The nitrogen abundance was determined by modeling \ion{N}{i} lines located around 7440 - 7480~\AA{} and 8600 - 8730~\AA{}. In Fig.~\ref{fig:R40_MOOG_Abundances} we can see our best model fit achieved for a nitrogen abundance of $\log\varepsilon{\rm (N)} = 8.55$, corresponding to [N/Fe]$=+1.25$ for the 2014 spectrum. It is important to cite that the lines in the 2014 spectrum display prominent asymmetric blue wings that are not well fitted. Such asymmetries may originate from the expansion of the pseudo-photosphere.

From the 2005 spectrum, our best fit model provides a nitrogen abundance of $\log\varepsilon{\rm (N)} = 8.65$, corresponding to [N/Fe]$=+1.35$ (Fig.~\ref{fig:R40_MOOG_Abundances}), assuming $T_\mathrm{eff}=9000$\,K, $\log{g}=1.5$.

Concerning the oxygen abundance, we fitted the lines at 6155.97, 6156.95, and 6158.17\,\AA{} in the spectrum taken in 2014 and obtained $\log\varepsilon{\rm (O)} = 8.0$, representing [O/Fe]$=-0.2$. This abundance is higher than derived by  \citet{1997ASPC..120...95V}, who estimated $\log\varepsilon{\rm (O)} = 7.7$ from spectra observed close to the maximum of the first (and hotter) eruption.

We also estimated the carbon abundance using the lines at 7465.45, 7470.09, 7473.31, and 7476\,\AA, resulting in $\log\varepsilon{\rm (C)} \leq 7.4$, equivalent to  [C/Fe]$\leq-0.5$. The low carbon abundance does not allow us to determine the carbon isotopic ratio of $^{12}$C/$^{13}$C, which is an important tool to identify evolved stars \citep{2009A&A...494..253K}.

The spectra of R\,40 during the eruption also show lines from s- and r-process elements, such as \ion{Ba}{ii}, \ion{La}{ii}, and \ion{Eu}{ii}. We calculated the europium abundance fitting the \ion{Eu}{ii} line at $6645$~\AA{} (Fig.~\ref{fig:R40_MOOG_Abundances}c), as $\log\varepsilon{\rm (Eu)} = 0.90$, corresponding to [Eu/Fe]$=+1.0$. For this modeling, the hyperfine splitting was taken from \citet{2008A&A...484..841M}. The lanthanum abundance was determined as $\log\varepsilon{\rm (La)} = 1.95$ ([La/Fe]$=+1.4$) from the 2014 spectrum using the line at 7483~\AA{}. We performed the same calculation to derive the barium abundance, using the \ion{Ba}{ii} line at 6142~\AA{} (Fig.~\ref{fig:R40_MOOG_Abundances}d), including the hyperfine and isotopic splitting taken from \citet{1998AJ....115.1640M}. Our best model corresponds to a barium abundance of $\log\varepsilon{\rm (Ba)} = 2.40$ ([Ba/Fe]$=+0.9$). This \ion{Ba}{ii} line is very strong and can be affected by NLTE effects, thus, our result has to be taken with caution. However, even considering the uncertainties of our modeling, the enrichment of s- and -r process elements seems to be real.

The projected rotational velocity, $v\,\sin i$, of R\,40 was also derived from the comparison of observed and synthetic spectra. Analyzing several metal lines that are present when the star is in the eruptive phase, we derived $v\sin i=23\pm 2$~km\,s$^{-1}$. On the other hand, during the quiescent phase, the modeling of the \ion{Mg}{ii} line provides $v\sin i=36\pm 3$~km\,s$^{-1}$. However, it is well known that atmospheres of supergiant stars are affected by strong macroturbulent motion, which was not considered in our synthetic spectra calculations. Thus, the values of 23 and 36~km\,s$^{-1}$ have to be considered as upper limits of the projected rotational velocity during the different phases.

\subsubsection{Discussion about R\,40}\label{sec:r40_discuss}

As already mentioned, this new eruption of R\,40 that started in 2007 is probably stronger than the previous eruption in 1988, based on its higher maximum brightness. The comparison of our 2014-2016 spectra, which correspond to a F8 supergiant star, with the spectrum reported by \citet{1993A&A...280..508S} as an A4-type, confirms a much cooler temperature during the second eruption. It is important to mention that the spectrum from \citet{1993A&A...280..508S} was not taken during the maximum of the first eruption, but probably not so far from that.

In addition, the estimated effective temperature from our modeling with {\sc moog} of $T_{\mathrm{eff,2015}} = 6500$~K is almost 1300~K lower than the value estimated by \citet{1993A&A...280..508S} for the first eruption. However, the apparent F8Iab type of our spectra indicates an even cooler temperature, around 6200~K for Magellanic Clouds supergiants, based on \citet{1984ApJ...284..565H} relation, which is also in agreement with our results using photometric and spectroscopic empirical calibrations. Concerning $\log{g}$, the value of $0.5$ adopted by our {\sc moog} modeling is the lower limit available in the Kurucz grid of equilibrium stellar models, and it is 0.25 dex lower than the value derived for the first eruption \citep{1993A&A...280..508S}. However, the true value can be even lower, since no hydrostatic and ionization equilibrium is present in a pseudo-photosphere.

Thus, even considering the effective temperature estimated by {\sc moog}, it is clear that R\,40 became one of the coolest identified LBVs with a temperature similar to that reported for the ongoing great eruption of R\,71 \citep{2013A&A...555A.116M}. It is also important to cite the similarities among the eruptions of both stars, indicating that R\,40 is probably facing its first observed great eruption.

Assuming the color excess of $E(B-V)=0.11\pm0.05$~mag \citep{1993A&A...280..508S, 2011ApJ...737..103S} and the empirical $R_V = A_V / E(B-V)= 2.74$ measured by \citet{2003ApJ...594..279G} for SMC, we could estimate a visual extinction of $A_V = 0.3 \pm 0.1$~mag. Thus, assuming the SMC distance modulus of $DM_{SMC} = 18.90$~mag \citep{1994MNRAS.266..441L}, the current mass of $M_{\mathrm{R\,40}} = 16\,M_\odot$ and the bolometric corrections ($BC$) from \citet{1984ApJ...284..565H}, we estimated the bolometric magnitude ($M_{\mathrm{Bol}}$), the effective radius of the pseudo-photosphere ($R$), and the $\log{g}$ for R\,40 in different epochs, as can be seen in Table~\ref{tab:R40_params}. 

The effective radius of R\,40 has increased by a factor of $\sim$5 since the star left its quiescence before 1985. Our results also point to $\log{g} = -0.1$ for the pseudo-photosphere, which is much lower than the value used by us in the {\sc moog} modeling. On the other hand, for the true quiescence of this star in 1960, we derived $\log{g_{1960}} = 1.25$, which is compatible with a blue supergiant star. For 1991, we derived $\log{g_{1991}} = 0.78$ and for 2005, $\log{g_{2005}} = 0.85$, indicating the star might not be in true quiescent phases because it was probably still affected by the material from previous eruptions.

\begin{table}
        \centering
        \caption{Bolometric magnitude ($M_{\mathrm{Bol}}$), effective radius ($R/R_\odot$), and $\log{g}$ for R\,40 in different epochs, based on the $V$ magnitude and bolometric corrections ($BC$) from \citet{1984ApJ...284..565H}. We assumed $A_V = 0.3 \pm 0.1$~mag for R\,40 and a distance modulus of $DM_{SMC} = 18.90$~mag for SMC. In the \textit{Year} column, Q means the quiescence and E the eruption stage.}
        \label{tab:R40_params}
        \small
        \begin{tabular}{ccccccc}
                \hline
                 Year  &   $V$   & Sp-type &  $BC$   & $M_{\mathrm{Bol}}$ & $R/R_\odot$ & $\log{g}$ \\ \hline\hline
                $1960(Q)$ & $10.73$ &  B8Ie   & $-0.51$ &   $-9.0 \pm 0.1$   &    $160$    &  $1.25$   \\
                $1991(E)$ & $10.15$ &   A4    & $-0.1$  &   $-9.2 \pm 0.1$   &    $280$    &  $0.78$   \\
                $2005(Q/E)$ & $10.3$  &  A0-A2  & $-0.2$  &   $-9.1 \pm 0.1$   &    $250$    &  $0.85$   \\
                $2016(E)$ &  $9.2$  &  F8Iab   & $0.08$  &   $-9.8 \pm 0.1$   &    $750$    &  $-0.1$   \\ \hline
        \end{tabular}
\end{table}

Considering the temperature variation as the source of R\,40 brightening during the ongoing eruption, the expected $\Delta V$ ($V$ apparent magnitude variation) should be only the difference of the $BC$ between the two epochs (quiescence and eruption). In our case, the spectral type changing from B8I to F8I, represents a changing of $BC$ from $-0.51$ to $0.05$, giving us a $\Delta V_{(M_{\mathrm{Bol}}=\mathrm{const.})} = \Delta BC = 0.56$, which cannot be explained just by the observed $\Delta V = 1.4$. This means that an increase of bolometric luminosity of about $0.8$~mag seems to be necessary. However, this increase needs to be seen with caution owing to the uncertain circumstellar extinction and especially the uncertain brightness of R\,40 and a possible UV excess during the quiescence. It is also possible that we are underestimating $M_{\mathrm{Bol}}$ during the quiescence because the bolometric correction for LBVs may be different from normal supergiants.

\begin{figure}
        \centering
        \includegraphics[width=\linewidth]{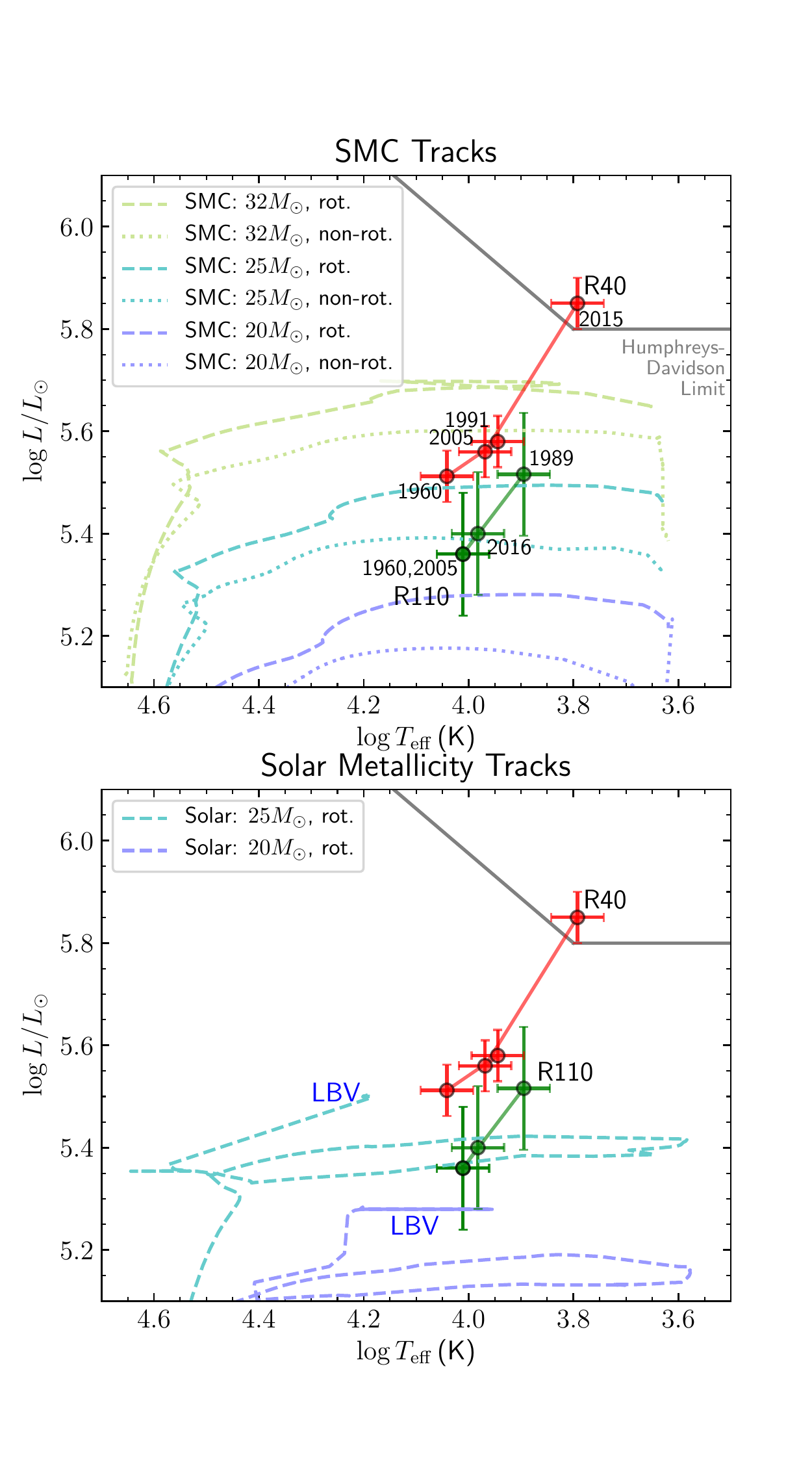}
        \caption{HR diagram showing the position of R\,40 and R\,110 in different epochs, based on our estimations. The Humphreys-Davidson limit is shown as a gray line. In addition, evolutionary tracks with rotation of $v_\mathrm{rot, ZAMS} = 0.4\,v_\mathrm{crit}$ (dashed lines) and without rotation (dotted lines) for SMC metallicity \citep[top panel]{2013A&A...558A.103G} and for solar metallicity \citep[bottom panel]{2012A&A...537A.146E} are also shown. In the bottom panel, we indicate the regions where LBV phase may occur according to \citet{2013A&A...550L...7G}.}
        \label{fig:hr_diagram}
\end{figure}

Considering the position of R\,40 in the HR diagram (Fig.~\ref{fig:hr_diagram}, top panel) associated with the evolutionary tracks with SMC metallicity from \citet{2013A&A...558A.103G}, we concluded that R\,40 is a star with $M_\mathrm{ZAMS} = 25 \sim 32\,M_\odot$. Its position on the evolutionary tracks points to a transition phase between BSG and RSG phases. However in these tracks, the LBV phase is not expected. On the other hand, considering tracks with solar metallicity from \citet{2012A&A...537A.146E}, shown in Fig.~\ref{fig:hr_diagram} (bottom panel), there is a blue loop after the RSG phase, allowing stars in this mass range, such as R\,40, to reach the LBV phase, owing to the mass loss during the RSG and yellow hypergiant (YHG) phase \citep{1994PASP..106.1025H, 2009ASPC..412...17O, 2013A&A...550L...7G}. In addition, R\,40 has crossed the Humphreys-Davidson limit \citep{1994PASP..106.1025H} during this ongoing eruption, confirming its nature as a strong eruption.

\begin{figure}
        \centering
        \includegraphics[width=\linewidth]{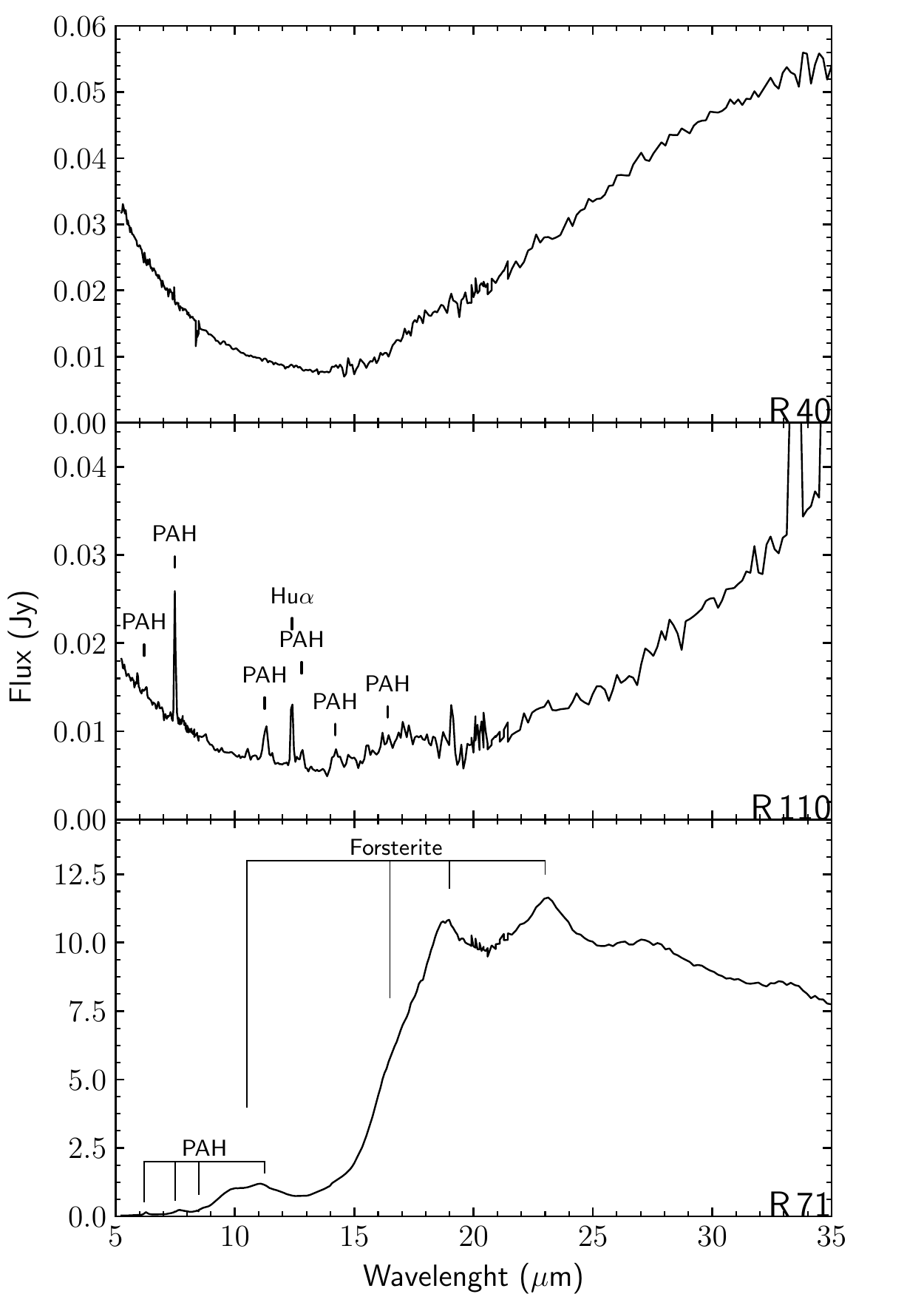}
        \caption{Spitzer IRS spectra of R\,40, R\,110, and R\,71. The spectra were obtained from the CASSIS public archive without any post-processing. The observation dates are 2008-08-04 for R\,40, 2004-07-18 for R\,110, and 2005-03-21 for R\,71. The identification of some spectral lines and dust features are provided.}
        \label{fig:spitzer}
\end{figure}

This post-RSG scenario for R\,40 would be supported not only by its low mass and luminosity, enrichment of nitrogen, and s- and r-process elements, but also by the probable presence of a dust shell, which is expected to be formed in the RSG phase \citep{1998Ap&SS.255..179W, 1999A&A...341L..67V, 1999PhDT.........3V, 2001ApJ...551..764L}. However, based on Spitzer IRS spectra, obtained from the Combined Atlas of Sources with Spitzer IRS Spectra (CASSIS) archive\footnote{\url{http://cassis.sirtf.com/}}, no strong silicate bands were observed. In Fig.~\ref{fig:spitzer}, we see the Spitzer IRS spectrum of R\,40 and also, for comparison, the spectrum of R\,71, where these bands are present, indicating the existence of a warm dust nebula \citep{1986A&A...164..435W, 1999A&A...341L..67V}. No PAH feature was identified in the IR spectrum of R\,40. Thus, we could not find any evidence of hot or warm dust around R\,40. On the other hand, the IR excess after 13~$\mu$m points to the possible existence of cold dust emission. This is in agreement with \citet{2013A&A...558A..17O} who studied the K-band spectrum of R\,40, identifying the presence of a double kinked continuum, as seen for YHGs, which may be associated with both a cool stellar photosphere and an infrared excess owing to cool dust. Thus, these authors suggested that R\,40 is an LBV in an eruptive stage, as confirmed by us.

Concerning the mass-loss rate during this current eruption, we used the relation between $\dot{M}$, $T_{\mathrm{eff}}$, $v_\mathrm{exp}$ (expansion velocity), and $L/L_{\odot}$ from \citet{1987ApJ...317..760D} (Eq. 4 and Fig. 1 in this reference), obtaining $\dot{M}_{2016} = 1 \times 10^{-4} \sim 1 \times 10^{-3}$~M$_\odot$\,yr$^{-1}$. For this rough estimate, we considered the same initial expansion velocity of $v_\mathrm{exp} = 10$~km\,s$^{-1}$, as assumed by \citet{2013A&A...555A.116M} for R\,71, which is a reasonable value for LBVs. In addition, as carried out by \citet{1980A&A....88...15W} for S Dor and by \citet{2013A&A...555A.116M} for R\,71, we also derive the mass-loss rate of R\,40 applying the equation $\dot{M} = 4\pi \mu m_H n_H v_\mathrm{exp} R^2$. Assuming the same density parameters ($\mu m_H n_H$) for the wind of R\,71 from \citet{2013A&A...555A.116M}, $R = 750 R_\odot$ and $v_{exp} = 10$~km\,s$^{-1}$, we obtain $\dot{M}_{R\,40} = 5 \times 10^{-4}$~M$_\odot$\,yr$^{-1}$. These two determinations have large uncertainties and we should consider these as just a rough estimation of the mass-loss rate. However, if such high values are really representative, we are observing an increase in the mass-loss rate of about 100 times for this ongoing eruption compared to the value derived by \citet{1993A&A...280..508S} for the quiescence of $\dot{M}_{1991} = 8 \times 10^{-6}$~M$_\odot$\,yr$^{-1}$.

\subsection{R\,110}\label{sec:R110}

\subsubsection{Light curve}

In Fig.~\ref{fig:light_curves}, we can see the maximum of the first detected eruption of R\,110, reported by \citet{1990A&A...228..379S}, which reached $V \sim 9.7$ in 1993. We discovered the presence of a second eruption, starting in 2005, but with uncertain maximum brightness owing to the lack of observations. Using AAVSO Visual band data, which is very similar to $V$ band, we obtained a maximum brightness of $V=9.9 \pm 0.2$~mag around 2011, which seems to be the peak of this second eruption. This maximum is similar to that of the first reported eruption. 

From data obtained in 2016, we can see the visual brightness of R\,110 is decreasing, probably returning to the quiescent stage. The minimum between these two eruptions seems to have reached $V \sim 10.9$ in 2005-2006, which was fainter than $V=10.5$ of the first measurements in $\sim$ 1981 \citep{1990A&A...228..379S} (see Fig.~\ref{fig:light_curves}).

\subsubsection{Spectral characteristics}

The spectral variability of R\,110 between 2005 and 2016 is less intense than that observed for R\,40 (see Fig.~\ref{fig:R110_B1}). The spectrum in this period is similar to an LBV in a quiescence, i.e., similar to a late-B or early-A spectral type with strong P~Cygni profiles in the Balmer lines and lines of singly ionized metals, such as \ion{Fe}{ii}, \ion{Cr}{ii,} and \ion{Ti}{ii}. We do not have any spectra for R\,110 close to the maximum of this current eruption around 2011 and our analysis may not reflect all the spectral variations due to this new eruption.

The presence of forbidden lines, mainly [\ion{Fe}{ii}] and [\ion{N}{ii}], with approximately symmetric line emission profiles, can be used to estimate the radial velocity of R\,110, $v_\mathrm{rad} = 265 \pm 5$~km\,s$^{-1}$. These lines are more intense in the spectrum taken in 2005, indicating the star was in a quiescent period with no signal of a previous pseudo-photosphere. After that, these lines became weaker, indicating that a new cool pseudo-photosphere was formed.

Similarly to R\,40, the Spitzer IRS spectra (Fig.~\ref{fig:spitzer}) of R\,110 shows no silicate bands. There is also an IR excess after 13\,$\mu$m, which is less intense than that observed in R\,40 and may also point to the presence of cool dust. We identified some possible PAH bands in the region from 6 to 17~$\mu$m, as shown in Fig.~\ref{fig:spitzer}. \citet{1999PhDT.........3V} cited that PAH molecules can be formed in oxygen-rich environments of LBVs owing to either the destruction of CO molecules by the UV radiation from the star or the shock of the current faster wind with the slower wind of a possible previous RSG phase. The possible band at 16\,$\mu$m, and also the strong peaks at 33 and 35\,$\mu$m seem to be due to reduction problems.

\subsubsection{Stellar parameters of R\,110}

From the \ion{Mg}{ii}/\ion{He}{i} equivalent widths ratio, we could estimate the spectral type and effective temperature of R\,110 (see Table~\ref{tab:R40_colortemperature}). In 2005, the star reached its highest temperature, $T_{\mathrm{eff}}\sim10500\,K$. Then, it dropped to about 9000~K in 2007 and 2014 and to about 8500\,K in 2016. This decrease of temperature from 2014 to 2016 is not expected based on the light curve behavior of R\,110, and it should be taken with caution owing to the possible wind emission contamination in the \ion{He}{i} lines.

We also used the same spectroscopic and photometric criteria as for R\,40 to estimate $T_\mathrm{eff}$ for R\,110. The results are also shown in Table~\ref{tab:R40_colortemperature}. 

\citet{2005A&A...439.1107D} used the equivalent widths of the \ion{Si}{ii} lines at $6374$~\AA{} and $6371$~\AA{} and of the \ion{He}{i} line at $6678$~\AA{} to determine the spectral type and temperature of LBVs, comparing the equivalent widths to measurements for standard Galactic stars. Using this relation, we found almost the same temperature for R\,110 in 2005, 2007, 2014, and 2016, which was on the order of $T_{\mathrm{eff}}= 10000 \pm 1000\,K$, resulting in a higher temperature during the eruptive phase than derived by other methods.

Since no \ion{Fe}{i} line was identified and \ion{Fe}{ii} lines have P~Cygni profiles (see Fig.~\ref{fig:S116_feii}), no modeling with {\sc moog} was possible for R\,110. 

Considering the extinction of R\,110, there are two different values for $E(B-V)$ reported in literature, that is, $E(B-V) = 0.44 \sim 0.75$ \citep{2011ApJ...737..103S} and $0.1 \sim 0.2$ \citep{1990A&A...228..379S}. The color excess from \citet{2011ApJ...737..103S} seems to be overestimated because this star is located in a region with inhomogeneous reddening \citep{2007ApJ...662..969I,2008A&A...484..205D}. Thus, the value from \citet{1990A&A...228..379S} was assumed by us to be $E(B-V) = 0.2 \pm 0.1$~mag.

Assuming the relation from \citet{1984ApJ...284..565H} as the source of bolometric correction, we estimated the bolometric magnitude ($BC$), effective radius ($R/R_\odot$), and $\log{g}$ for R\,110 in different dates. The results are shown in Table~\ref{tab:R110_params}, where it is possible to see a variability of these parameters during the second eruption, which is not as strong as seen for R\,40. For $\log{g}$ estimation, we assumed a current mass of $M_{\mathrm{R\,110}} = 10\,M_\odot$ \citep{1990A&A...228..379S}.

\begin{table}
        \centering
        \caption{Bolometric magnitude ($M_{\mathrm{Bol}}$), effective radius ($R/R_\odot$), and $\log{g}$ for R\,110 in various epochs based on the $V$ magnitude and bolometric corrections ($BC$) from \citet{1984ApJ...284..565H}. We assumed $A_V = 0.62 \pm 0.3$~mag for the star and a distance modulus of $DM_{LMC} = 18.50$~mag for the LMC. In the \textit{Year} column, Q means the quiescence, and E the eruption stage.}
        \label{tab:R110_params}
        \small
        \begin{tabular}{ccccccc}
                \hline
                 Year  &  $V$   & Sp-type &  $BC$   & $M_{\mathrm{Bol}}$ & $R/R_\odot$ & $\log{g}$ \\ \hline\hline
                $1960(Q)$ & $10.9$ &  B9Ieq  & $-0.38$ &   $-8.6 \pm 0.3$   &    $150$    &  $1.1$   \\
                $1989(E)$ & $9.99$ &   F0    & $-0.1$  &   $-9.0 \pm 0.3$   &    $310$    &  $0.45$   \\
                $2005(Q)$ & $10.9$ &   B9    & $-0.2$  &   $-8.6 \pm 0.3$   &    $150$    &  $1.1$   \\
                $2016(E)$ & $10.5$ &   A2    & $0.08$  &   $-8.7 \pm 0.1$   &    $200$    &  $0.83$   \\ \hline
        \end{tabular}
\end{table}

\subsubsection{Discussion about R\,110}

In contrast to what was seen for R\,40, R\,110 is nowadays experiencing a weaker eruption than the first eruption reported by \citet{1989AGAb....3..115S}. The analysis of its light curve indicates that the maximum intensity in the $V$ band of this new eruption is about $0.3$ magnitude weaker than in the first eruption, reaching $V=9.9 \pm 0.2$~mag around 2011. However, this maximum is not well determined owing to the lack of data between 2011 and 2015.

Based on the spectral variability, \citet{1989AGAb....3..115S} reported an F spectral type in 1989 for R\,110. On the other hand, based on our analysis, it remains an early A star from 2005 until 2016. None of our estimates point to a temperature lower than $T_{\mathrm{eff}} = 8500$~K for the current eruption, which is higher than $T_{\mathrm{eff}} = 7600$~K reported by \citet{1989AGAb....3..115S} for the first eruption. It is important to note that our analysis is incomplete because we do not have spectroscopic data close to the probable maximum of the eruption in 2011.

Between these two eruptions, the star stayed for a short time in a true quiescence between 2003 and 2004, when it reached its minimum and presented a B9 spectral type (Table~\ref{tab:R40_colortemperature}) and $V=10.9$~mag, similar to the values described by \citet{1960MNRAS.121..337F} before the first eruption. This spectral type corresponds to $T_{\mathrm{eff}}\sim10500$~K for stars with LMC metallicity. 

Based on the light curve, we can also see that during the first observed eruption, the star showed a brightness increase during more than 10 years ($\sim$ 1982-1993) and a similar time period for its brightness decline ($\sim$ 1994-2003). On the other hand, the new eruption showed a much shorter brightness growth, going from the quiescence to the maximum of the eruption in around 5-7 years. This may indicate a higher mass-loss rate during a shorter period for this new eruption compared to the first eruption. 

From our analysis for R\,110, we found an almost constant $M_{\mathrm{Bol}} = -8.8 \pm 0.2$~mag, assuming the reddening from \citet{1990A&A...228..379S}. This value is in agreement with the literature and it includes R\,110 in the group of low luminosity LBVs, which may have previously passed through a RSG phase.

The $T_\mathrm{eff}$ and $M_\mathrm{Bol}$ estimated for R\,110 points to a location in the HR diagram in which the evolutionary tracks from \citet{2013A&A...558A.103G} indicate a transition phase from BSG to RSG for stars with $20 \leq M_\mathrm{ZAMS} \leq 25\,M_\odot$. However, as was also cited for R\,40, these tracks do not predict the existence of the LBV phase.

\begin{figure*}
        \centering
        \includegraphics[width=\linewidth]{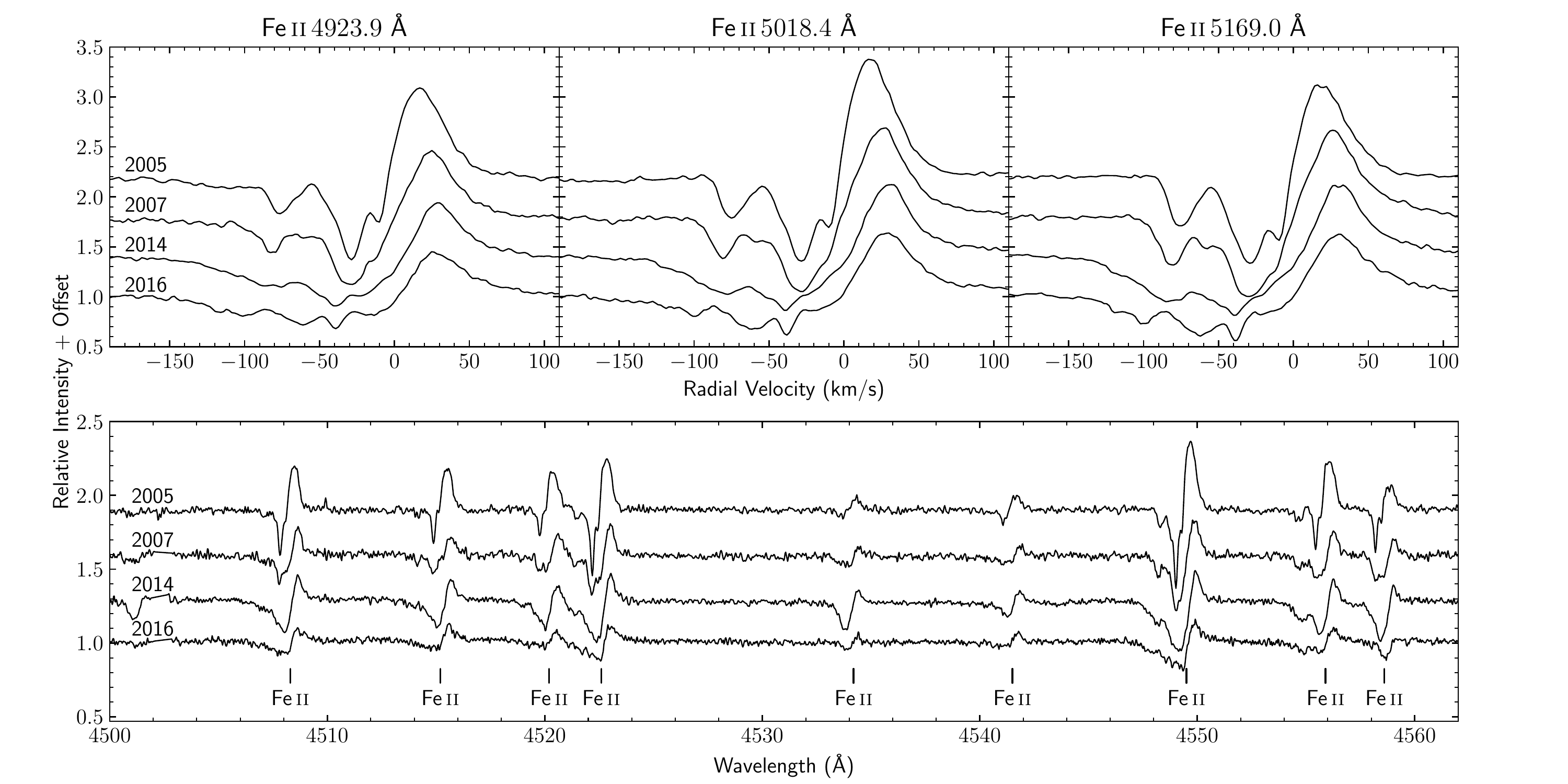}
        \caption{\ion{Fe}{ii} lines of R\,110 observed in 2005, 2007, 2014, and 2016. In the top panel, we have \ion{Fe}{ii} lines in 4923.9, 5018.4, and 5169.0~\AA{} (from the multiplet 42), which are the strongest lines of this ion in the spectrum of R\,110 and present P~Cygni profiles with multiple variable absorption components. The bottom panel shows the spectral region from 4500 to 4565~\AA{} with other \ion{Fe}{ii} lines with P~Cygni profiles.}
        \label{fig:S116_feii}
\end{figure*}

The most remarkable point in the spectral analysis of R\,110 is the variability of radial velocity related to P~Cygni multiple absorption components seen in \ion{Fe}{ii} and Balmer lines (see Fig.~\ref{fig:S116_feii} and ~\ref{fig:R110_balmer}, respectively). The appearance of these multiple components is possibly due to different shells in the expanding wind, which are formed by an increase of the mass-loss rate during the eruption or, as in the case of \object{AG~Car} (a Galactic LBV), by dramatic changes in the terminal velocity and mass-loss rate owing to the bi-stability jump effect \citep{2009ApJ...698.1698G}. 

\section{Conclusions}\label{sec:conc}

We have identified new ongoing eruptions for two LBVs in the Magellanic Clouds, R\,40 in SMC and R\,110 in LMC. These two new eruptions are occurring in very different scenarios: R\,40 is experiencing a strong eruption that is much stronger than the previous eruption observed in 1996; and R\,110 is now passing through a weak eruption that is weaker than the previous eruption in 1994. Our results confirm the suggestion made by \citet{2013A&A...558A..17O}, based on K-band spectra, that both stars should be suffering eruptions. 

From our study, based on new spectroscopic and photometric data obtained by us, and supplemented by data from the literature, we derived physical parameters for both stars in different phases: quiescence and eruption.

R\,40, due to its new eruption, became one of the coolest LBVs already observed with $T_{\mathrm{eff}} = 5800-6500$~K. During the quiescent period between the last two eruptions, R\,40 still showed evidence of the effects from the first eruption: absence of forbidden lines, a lower visual magnitude, and a lower effective temperature than during the quiescent phase previous to the first reported eruption. From the modeling of observed spectra during the ongoing eruption, we identify an enrichment of nitrogen and r-process elements. These results, associated with the possible presence of a cool dust shell, indicate a post-RSG nature for this object. An enrichment of Ba was also derived, which is not seen in other post-RSG stars \citep{2016MNRAS.461.4071S}.

R\,110 had, between 2003 and 2004, a true quiescence with the presence of forbidden lines, strong P~Cygni profiles, and maximum visual magnitude, keeping the same spectral type compared to the previous quiescence, as reported by \citet{1960MNRAS.121..337F}. It is most likely that during the quiescence, R\,110 had an effective temperature that was not higher than $10500$~K. On the other hand, because of the new eruption, the temperature dropped to not lower than $8500$~K. Based on its low luminosity and temperature, we cannot discard a post-RSG nature for R\,110 either.

Based on our estimated parameters for both stars in different epochs, we can see in Fig.~\ref{fig:hr_diagram} how R\,40 and R\,110 are evolving from the quiescence to eruption and vice versa in the HR diagram. As for each star just two eruptions have been recorded, it is not possible to derive a periodicity for these events. Thus, an observational campaign, associated with photometry, spectro-polarimetry and high-resolution spectroscopy, covering not only the V band but also other bands is definitely necessary to follow up these eruptions and better derive their characteristics, and also identify new eruptions. A deeper abundance study is also necessary to confirm a previous RSG phase for these low luminosity LBVs. 

\begin{acknowledgements}
We thank the anonymous referee for comments that helped us to improve the paper.

J. C. N. C. acknowledges Coordenadoria de Aperfei\c{c}oamento de Pessoal de N\'{i}vel Superior (Capes) for the PhD grant.

N.A.D. acknowledges the St. Petersburg State University for the research grant 6.28.335.2015 and FAPERJ, Rio de Janeiro, Brasil, for the visiting researcher grant E-25/200.128/2015.

M.K. acknowledges financial support from GA\v{C}R (grant number 17-02337S). The Astronomical Institute Ond\v{r}ejov is supported by the project RVO:67985815. This research was also supported by the European Union European Regional Development Fund, project ``Benefits for Estonian Society from Space Research and Application'' (KOMEET, 2014\,-\,2020.\,4.\,01.\,16\,-\,0029) and by the institutional research funding IUT40-1 of the Estonian Ministry of Education and Research. Parts of the observations obtained with the MPG 2.2 m telescope were supported by the Ministry of Education, Youth and Sports project - LG14013 (Tycho Brahe: Supporting Ground-based Astronomical Observations). We would like to thank the observers (S. Ehlerova, P. Kabath, A. Kawka) for obtaining the data.

C. A. G. acknowledges financial support through a \textit{Nota 10} fellowship granted by FAPERJ (Funda\c{c}\~ao Carlos Chagas Filho de Amparo \`{a} Pesquisa do Estado do Rio de Janeiro).
\end{acknowledgements}

\bibliographystyle{aa}
\bibliography{referencias}{}

\onecolumn

\Online

\begin{appendix} 
\section{Spectral variability of R\,40}

\begin{figure}[h]
        \centering
    \includegraphics[width=\linewidth]{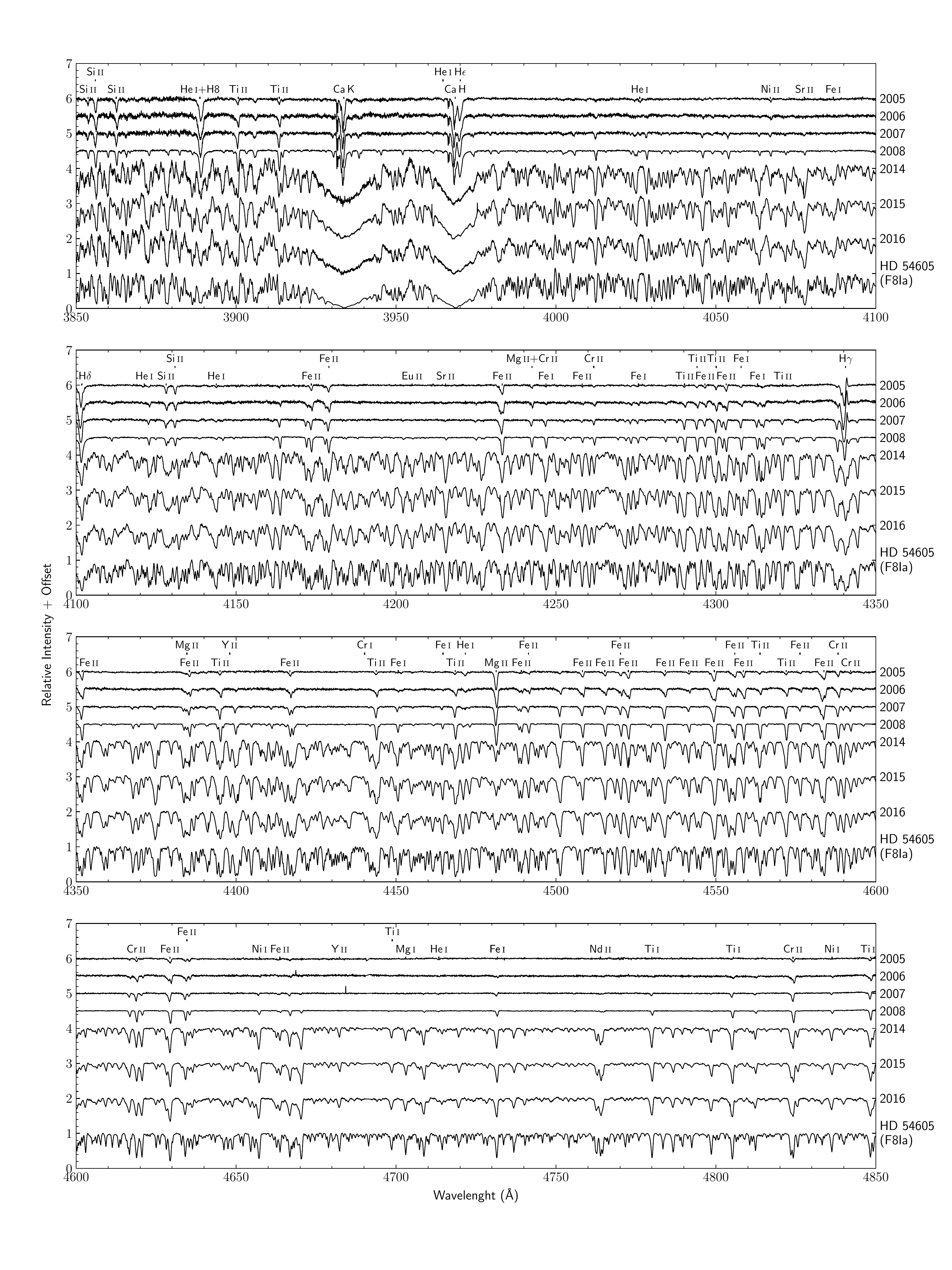}
        \caption{Complete optical spectra of R\,40 taken from 2005 until 2016 with FEROS. The spectra are radial velocity corrected, but without telluric correction. It is possible to see the changing from a late B-type or early A-type to a late-F spectrum owing to the newly identified eruption. The spectrum of HD~54605 (F8Iab) is also shown for comparison. The identification of the lines, based on \citet{1945MOORE}, \citet{1996ApJS..103..183L}, \citet{1997AJ....114..376C}, and \citet{2002A&A...387..560H} are also provided.}
        \label{fig:A1}
\end{figure}

\begin{figure}[h]
    \centering
    \ContinuedFloat
    \includegraphics[width=\linewidth]{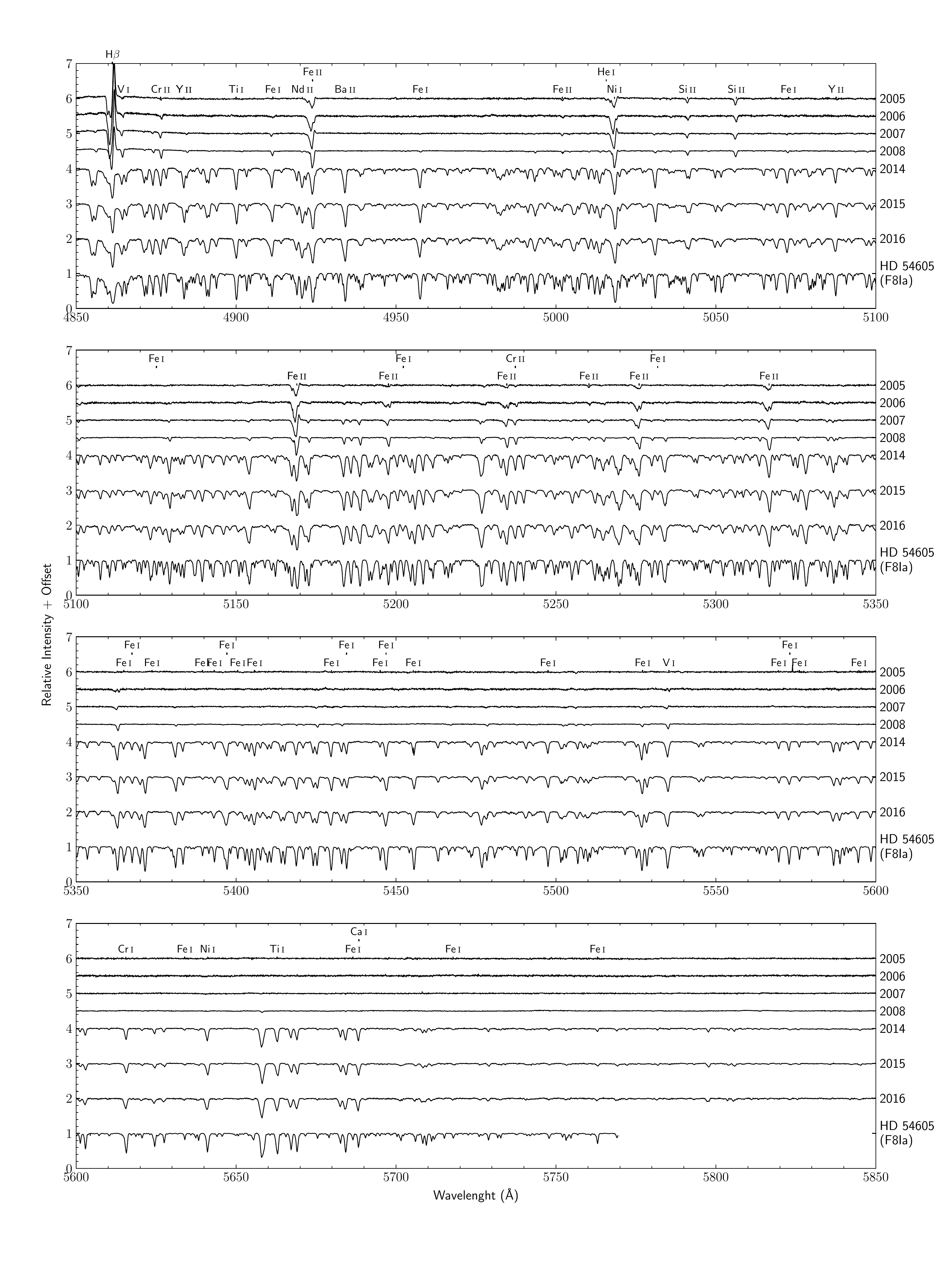}
    \caption{continued.}
\end{figure}

\begin{figure}[h]
    \centering
    \ContinuedFloat
    \includegraphics[width=\linewidth]{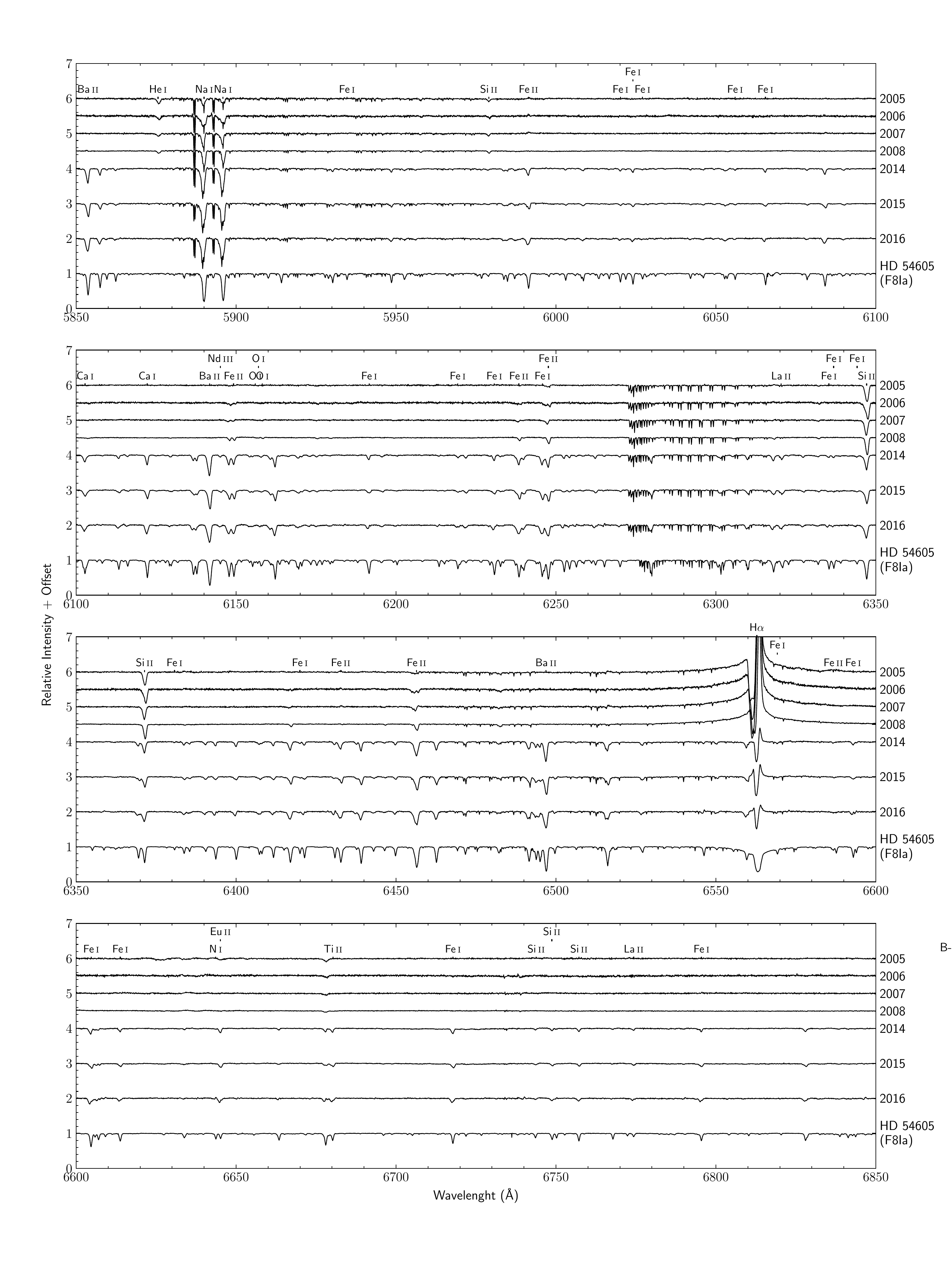}
    \caption{continued.}
\end{figure}

\begin{figure}[h]
    \centering
    \ContinuedFloat
    \includegraphics[width=\linewidth]{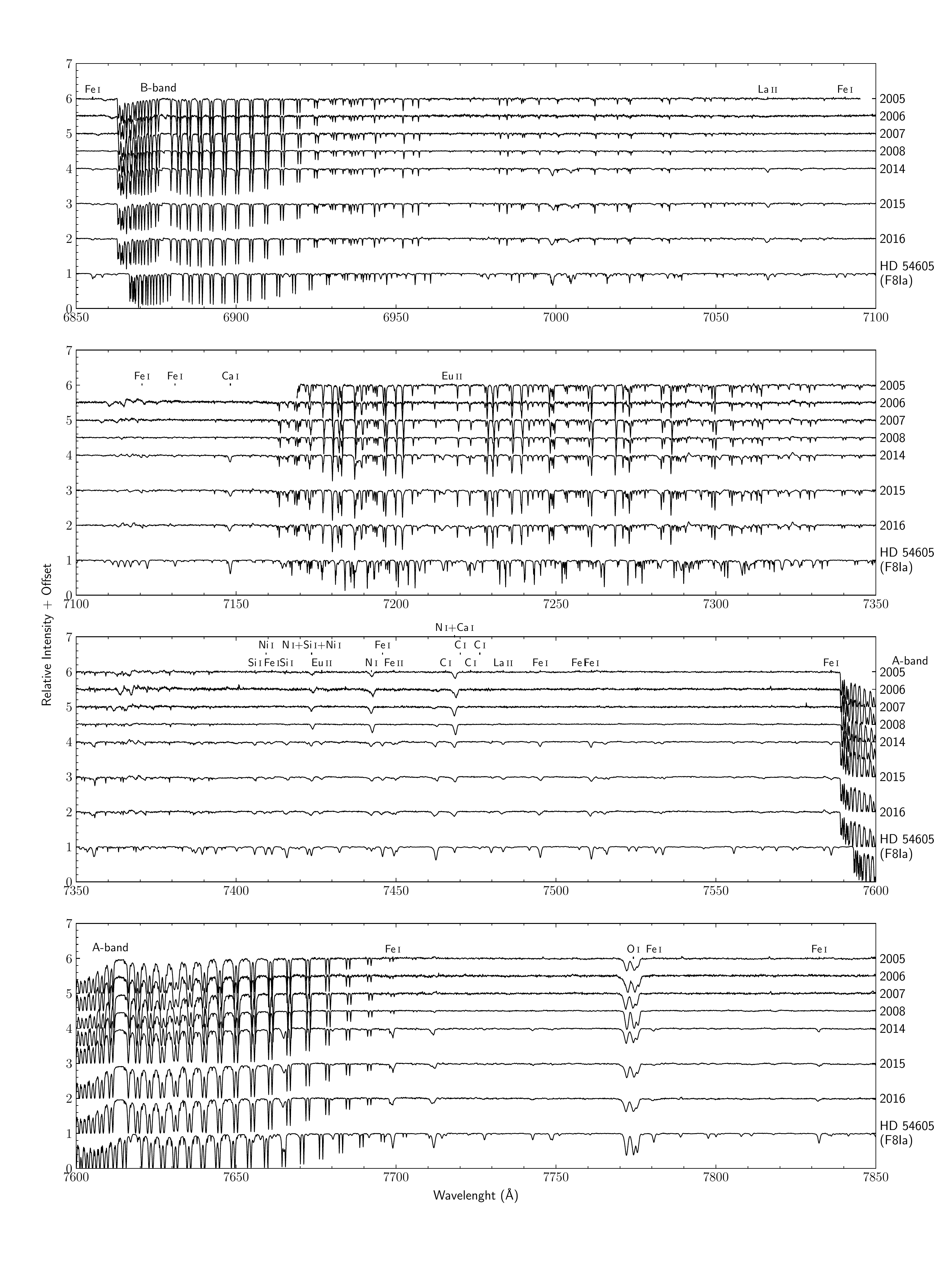}
    \caption{continued.}
\end{figure}

\begin{figure}[h]
    \centering
    \ContinuedFloat
    \includegraphics[width=\linewidth]{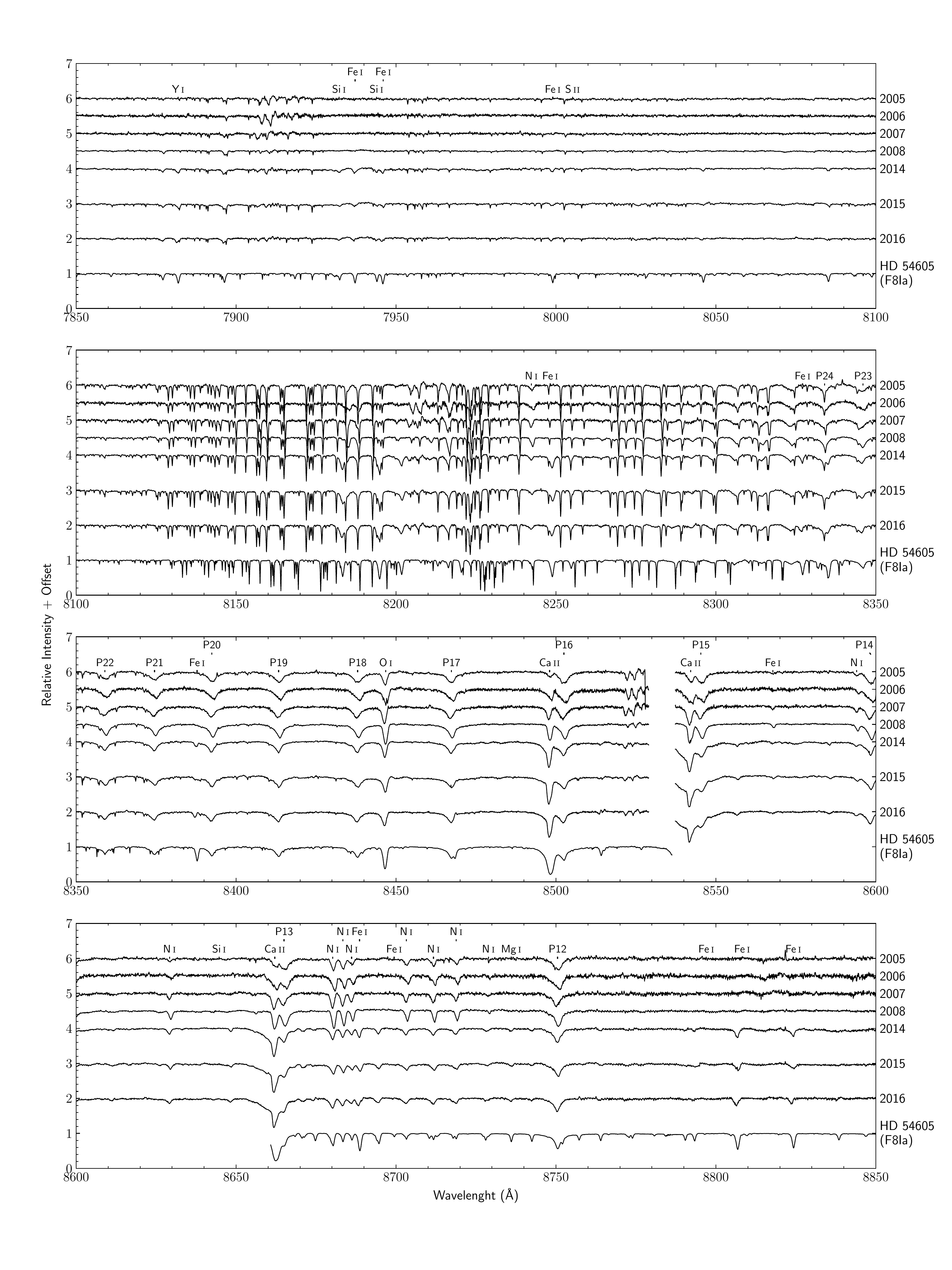}
    \caption{continued.}
\end{figure}

\begin{figure}
        \centering
        \includegraphics[width=\linewidth]{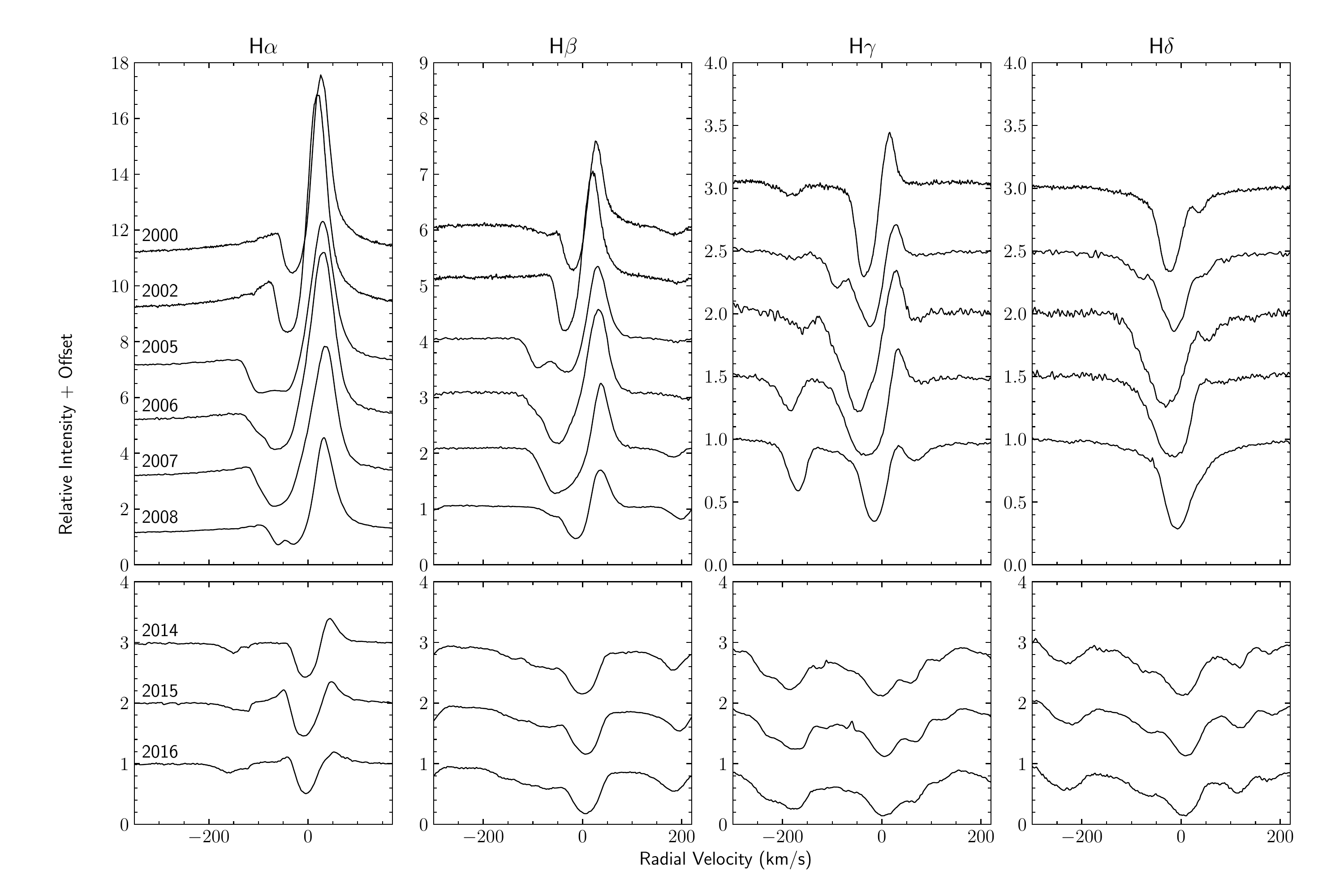}
        \caption{Variation of H$\alpha$, H$\beta$, H$\gamma,$ and H$\delta$ line profiles of R\,40 shown in various dates.}
        \label{fig:S52_Halpha}
\end{figure}

\section{Spectral variability of R\,110}

\begin{figure}[h]
        \centering
    \includegraphics[width=\linewidth]{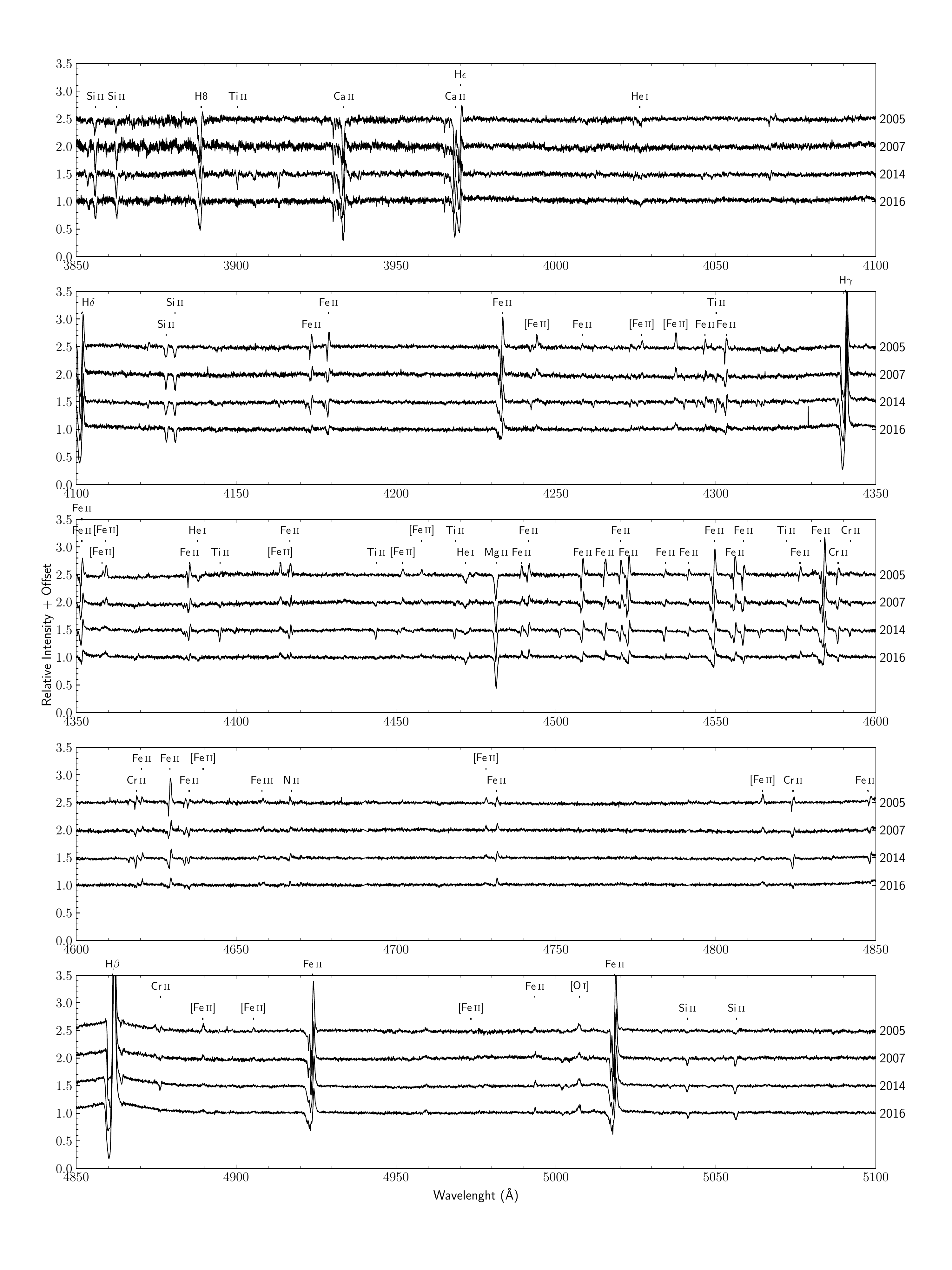}
        \caption{Complete optical spectra of R\,110 taken from 2005 until 2016 with FEROS. The spectra are radial velocity corrected, but without telluric correction. It is possible to see the variations caused by the newly identified eruption. The identification of the lines, based on \citet{1945MOORE}, are also provided.}
        \label{fig:R110_B1}
\end{figure}

\begin{figure}[h]
    \centering
    \ContinuedFloat
    \includegraphics[width=\linewidth]{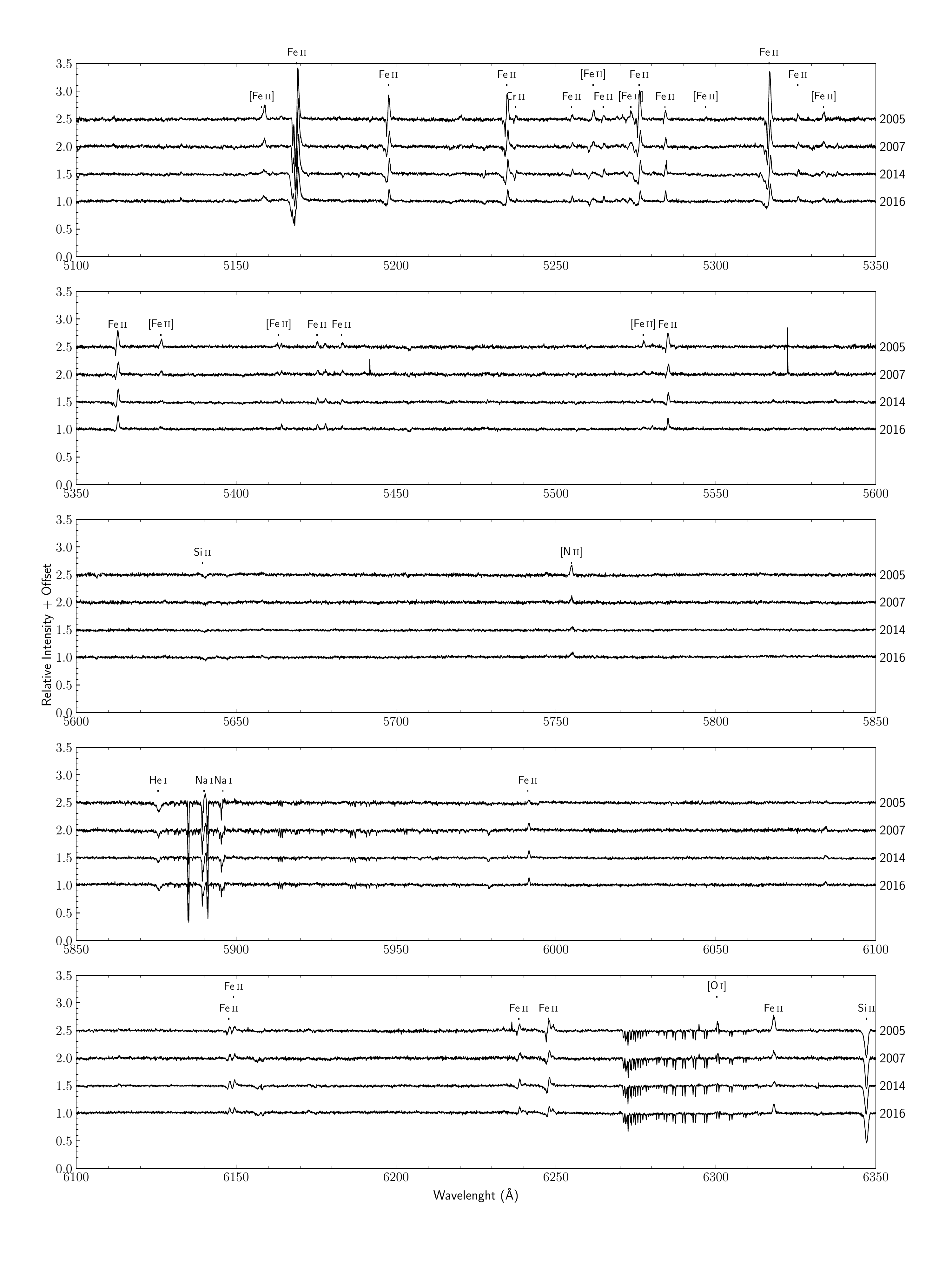}
    \caption{continued.}
\end{figure}

\begin{figure}[h]
    \centering
    \ContinuedFloat
    \includegraphics[width=\linewidth]{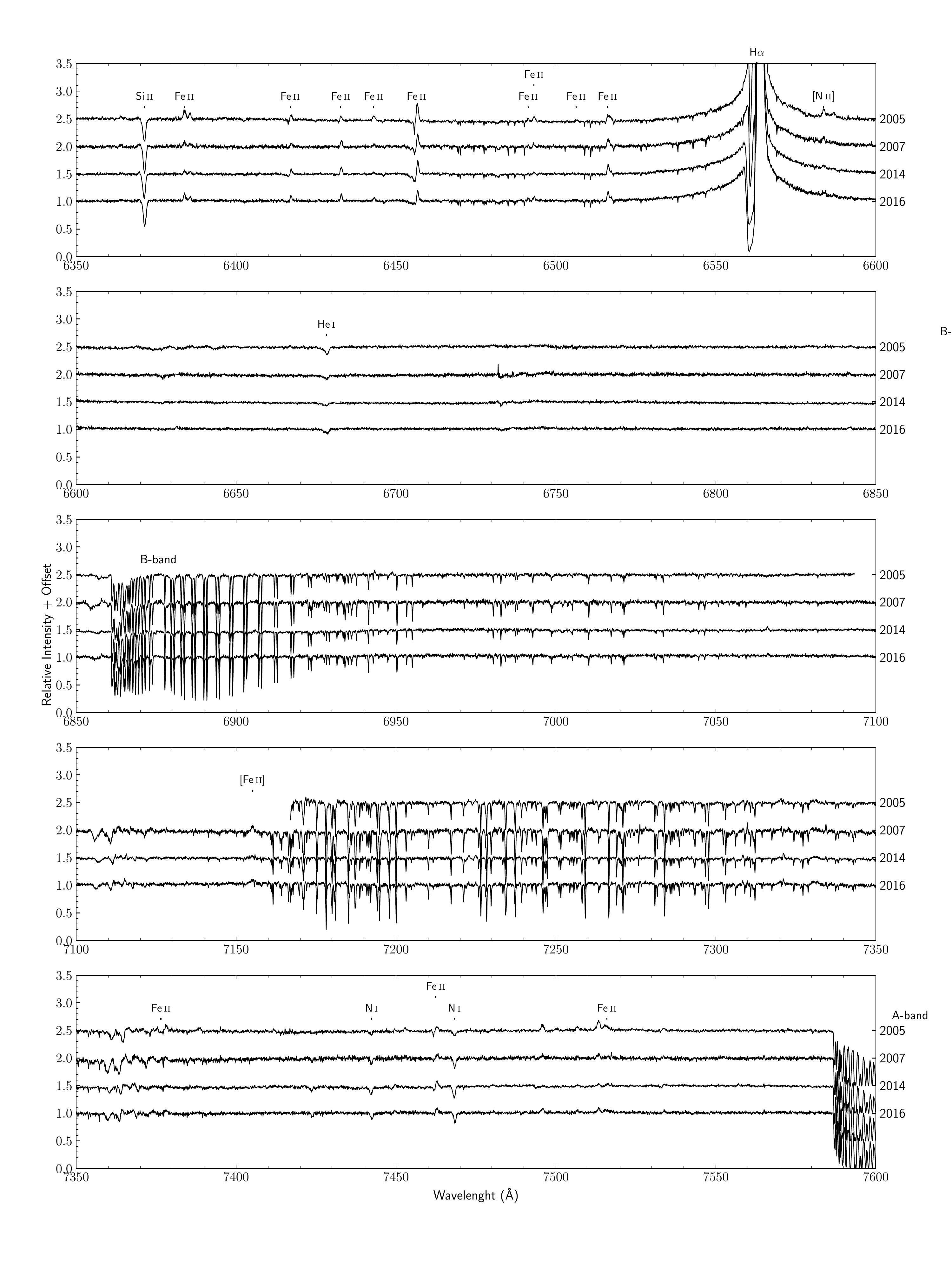}
    \caption{continued.}
\end{figure}

\begin{figure}[h]
    \centering
    \centering
    \ContinuedFloat
    \includegraphics[width=\linewidth]{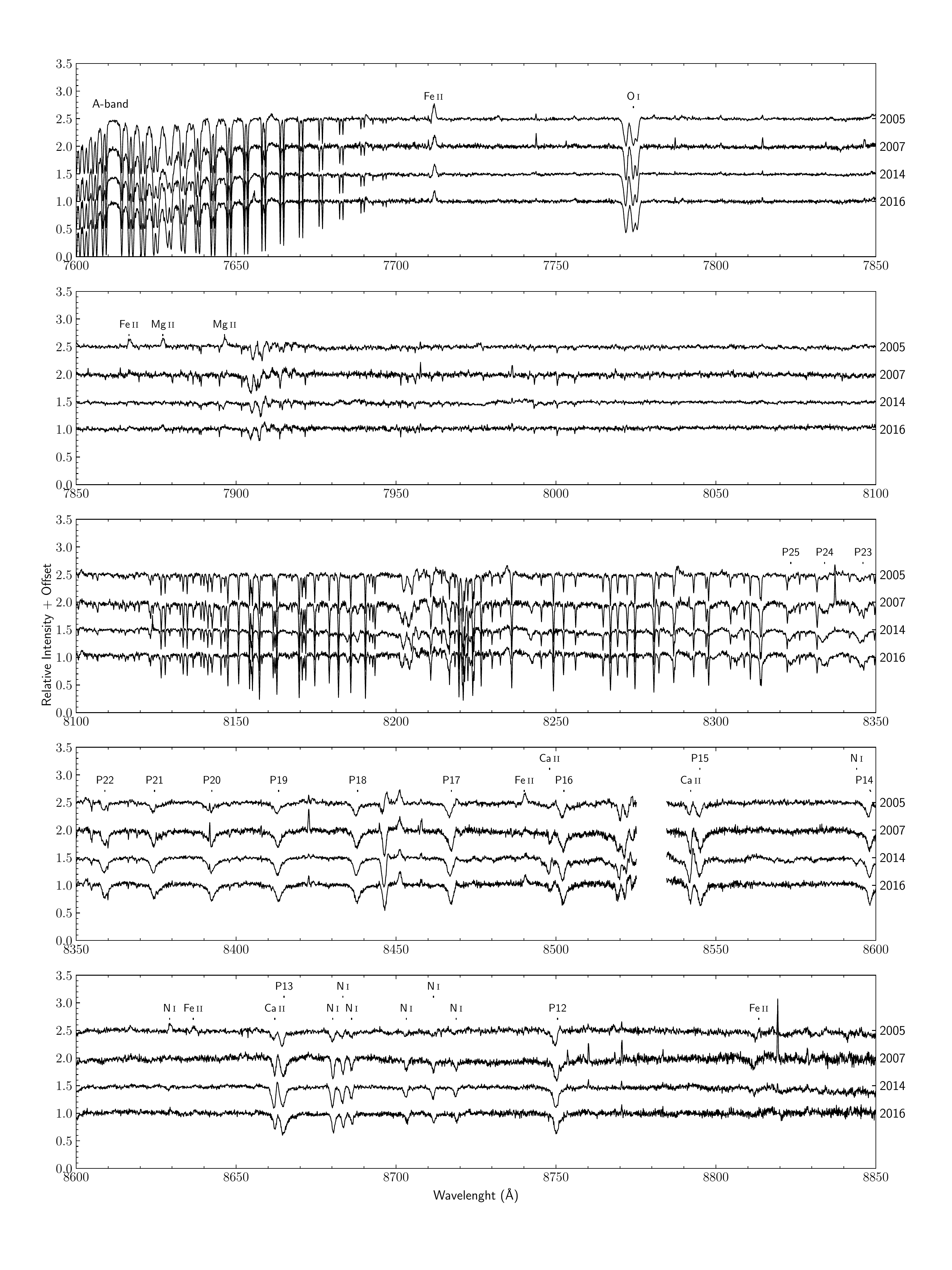}
    \caption{continued.}
\end{figure}

\begin{figure}
        \centering
        \includegraphics[width=\linewidth]{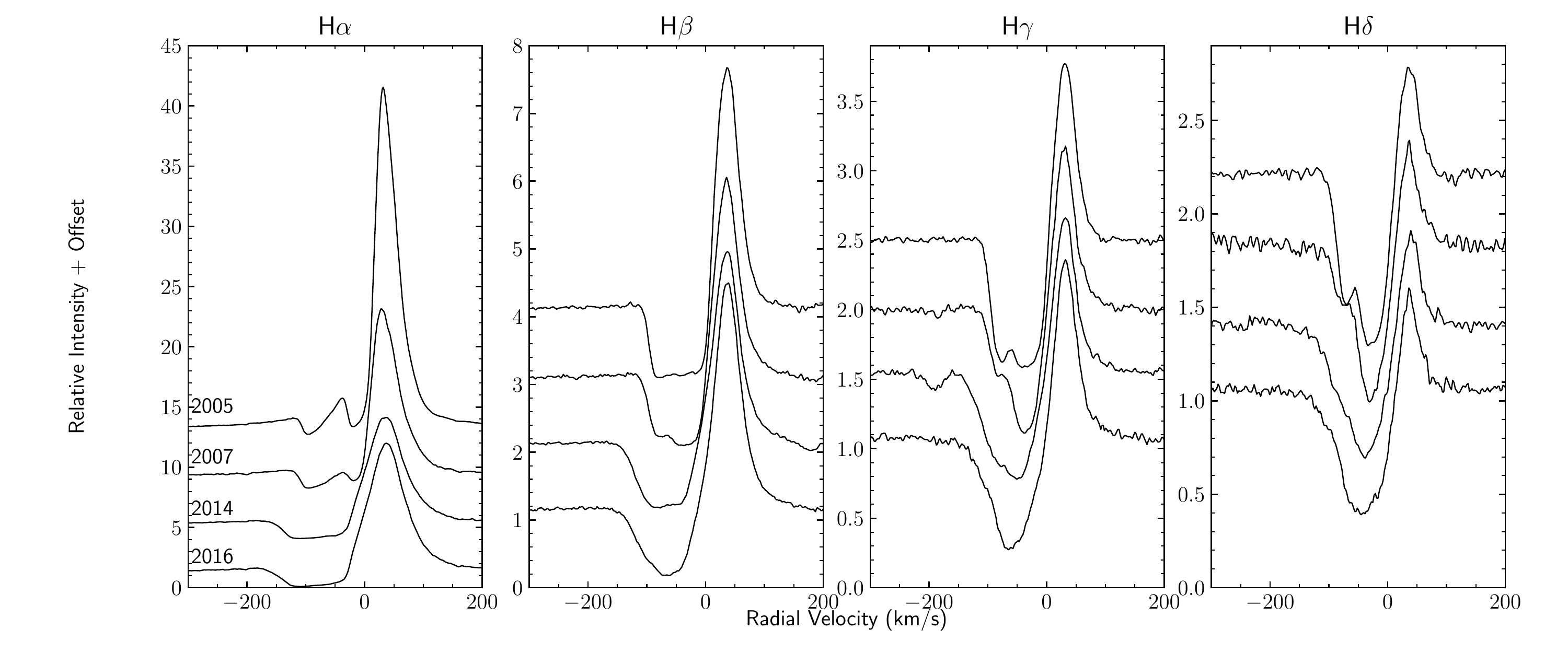}
        \caption{Balmer lines (H$\alpha$, H$\beta$, H$\gamma,$ and H$\delta$) of R\,110 observed in 2005, 2007, 2014, and 2016.}
        \label{fig:R110_balmer}
\end{figure}
\end{appendix}  

\end{document}